\documentclass[aps,prd,amsmath,amssymb]{revtex4}
\usepackage{graphicx}
\usepackage{bm}

\def\journal#1#2#3#4{{#1} {\bf #2}, #3 (#4)}
\newcommand{\be}{\begin{equation}}
\newcommand{\ee}{\end{equation}}
\newcommand{\bea}{\begin{eqnarray}}
\newcommand{\eea}{\end{eqnarray}}
\newcommand{\hf}{\frac12}
\newcommand{\nn}{\nonumber\\}
\def\eq#1{(\ref{#1})}
\def\la{\langle}
\def\ra{\rangle}

\def\Tr{{\mathrm{Tr}}}
\def\ord#1{{\cal O}\left(#1\right)}
\def\mr#1{{\mathrm{#1}}}
\def\v#1{{\bm{#1}}}
\def\fd#1#2{\frac{\delta#1}{\delta#2}}
\def\fdd#1#2#3{\frac{\delta^2#1}{\delta#2\delta#3}}
\def\dt{{\Delta t}}
\def\hj{{\hat j}}

\def\hx{\hat x}

\def\hD{{\hat D}}

\def\ih{\frac{i}{\hbar}}

\def\sign{\mr{sign}}

\begin{document}
\title{Environment Induced Time Arrow}
\author{Janos Polonyi}
\affiliation{Strasbourg University, High Energy Theory Group, CNRS-IPHC,
23 rue du Loess, BP28 67037 Strasbourg Cedex 2 France}
\date{\today}

\begin{abstract}
The spread of the time arrows from the environment to an observed subsystem is followed within a harmonic model. A similarity is pointed out between irreversibility and a phase with spontaneously broken symmetry. The causal structure of interaction might be lost in the irreversible case, as well. The Closed Time Path formalism is developed for classical systems and shown to handle the time arrow problem in a clear and flexible manner. The quantum case is considered, as well, and the common origin of irreversibility and decoherence is pointed out.
\end{abstract}
\maketitle
\tableofcontents

\section{Introduction}
According to our daily experience the time flows ``ahead`` and events look differently when seen in a movie played backward. The time arrow which points in the direction of the flow of time seems to be well defined despite the invariance of the fundamental equations of motion under time reversal, $t\to-t$, the weak interaction put aside. There are two possible levels where the orientation of time might be defined. One is General Relativity where both the time and its arrow are constructed. The other is in quantum mechanics or thermodynamics where quantum laws or statistical mechanics sets the time arrow \cite{zehta,savitt,mackey}. The issue of the mechanical time arrow is considered in this work by paying special attention to the preparation of the system and the way it is observed. The information we need about the preparation is encoded in the boundary conditions in time, imposed on the solution of the dynamical problem and what is assumed about the observations is that they are made in a finite duration of time.

The main idea, borrowed from the renormalization group method \cite{kadanoff,wilson}, is that in any real experience on a dynamical system, called full system below, we observe a subsystem only, called briefly system from now on, and can construct an effective theory for this subsystem. The effective theory can in principle be obtained from the detailed dynamics of the full system by eliminating the non-observed degrees of freedom, called environment, by solving their equations of motion. One can speak of irreversibility with respect to the observed subsystem only since it refers to the dynamical process where the environment boundary conditions induce a time arrow for the observed subsystem of the full system obeying time reversal invariant dynamics. It may happen that the environment induces interactions both forward and backward in time and no definite system time arrow is observed, the system dynamics is acausal in such cases. 

It seems to us that the emergence of irreversibility can be understood as a dynamical symmetry breaking. This view is motivated by a careful separation of two scenarios, the realistic one based on observations carried out in arbitrarily long but finite time and another, mathematically simpler idealized situation where observations may take infinitely long time. Though irreversibility and acausality share the same dynamical origin their manifestations are different. The former is a spontaneous symmetry breaking equipped with an order parameter and is easier to recognize. The observation of the latter is more involved which may explain the more abundant models and dynamical scenarios, developed to find the dynamical origins of irreversibility than those of acausality.

The Closed Time Path (CTP) formalism \cite{schw} is used in this work. It has already been reinvented in different quantum contexts such as to derive relaxation in many-body systems \cite{kadanoffb}, to develop perturbation expansion for retarded Green-functions \cite{keldysh}, to find manifestly time reversal invariant description of quantum mechanics \cite{aharonov}, to describe finite temperature effects in quantum field theory \cite{umezawa,niemisa,niemisb}, to find the mixed state contributions to the density matrix by path integral \cite{feynmanv}, to follow non-equilibrium processes \cite{calzetta}, to derive equations of motion for the expectation value of local operators \cite{jordan,ed}  and to describe scattering processes with non-equilibrium final states \cite{scatt}. This scheme, introduced here already in classical mechanics does not give rise to new equations. However it is better suited to the problem considered because it separates clearly the oriented and unoriented correlations in time among the dynamical degrees of freedom and sources, it can deal with initial condition problem for an infinite system within the framework of the action principle thereby allowing the use of powerful functional methods. Furthermore, it provides a variation principle to derive dissipative effective equations of motion and finally it makes clear that causality is not automatic for infinite systems. 

We start with a succinct description of our procedure in Section \ref{syn}, followed by the introduction of our model in Section \ref{clho}, the presentation of irreversibility as a spontaneous symmetry breaking in Section \ref{ssb} and the introduction of the CTP scheme for classical and quantum systems in Sections \ref{cctp}-\ref{qctp}, respectively. The summary of the results is presented in Section \ref{sum} and an Appendix contains the derivation of the CTP Green-function.

\section{Synopsis}\label{syn}
The issue of the time arrow is addressed in this work in a full system of $N+1$ degrees of freedom, obeying conservative, time reversal invariant dynamics described by the Lagrangian $L$. We choose a system coordinate, denoted by $x_0$, the remaining $N$ degrees of freedom making up the environment. One encounters two different system equations of motion. The Euler-Lagrange equation for $x_0$ derived from the Lagrangian $L$ is a non-autonomous differential equation. However once the environment coordinates are eliminated by means of their equations of motion, this equation becomes an effective, autonomous integro-differential equation. A well known case where it is important to distinguish these two kinds of equations is the radiation reaction force in classical electrodynamics \cite{rohrlich}, sought in vain in the elementary equations of motion. 

We follow in this work the change of the elementary equation into the effective one and identify the point where the time arrow is generated. For this end it is advantageous to separate two kinds of boundary conditions. Those needed to select a unique solution of the effective system equation of motion, called system boundary conditions which remain externally imposed parameters of the solution. The remaining, environment boundary conditions are build into the effective system equation of motion and are the dynamical source of irreversibility, a dynamically generated system time arrow. A necessary condition for irreversibility is to have infinitely many environment initial conditions.

A simple harmonic model is introduced and its classical dynamics is studied in {\bf Section \ref{clho} Classical harmonic systems}. The model, defined in {\bf Section \ref{sysoscs} System of oscillators} consists of linearly coupled harmonic oscillators \cite{rubin,senitzky,ford,ullersma,caldeira} and is describe by the help of the Lagrangian
\be\label{lagrho}
L=\frac{m}{2}\dot x_0^2-\frac{m\omega^2_0}2x_0^2+jx_0+\sum_{n=1}^N\left(\frac{m}2\dot x^2_n-\frac{m\omega_n^2}2x^2_n-g_nx_nx_0\right),
\ee
where a time dependent external source $j$ is coupled to the system coordinate $x_0$. The system trajectory $x_0(t)$, subject to the initial conditions $x_0(0)=\dot x_0(0)=0$ is constructed in this Section in terms of the external source $j(t)$ and the retarded Green-function of the system. 

It is important to distinguish two kinds of time arrow for each degree of freedom. If we impose initial or final conditions on a coordinate then the corresponding degree of freedom has a forward or a backward pointing internal time arrow. When boundary conditions are used in time, like for variational problems, then no internal time arrow exists. Either initial or final conditions are allowed in the model \eq{lagrho} and the retarded Green-function for the system coordinate $x_0(t)$ contains in its denominator a self energy term, written in terms of the retarded or advanced Green-functions of the environment oscillators with forward or backward oriented internal time arrow, respectively. The real, dynamical time arrow, called simply time arrow from now on is defined for each degrees of freedom by the direction in time in which a linearly coupled source generates a response. In case of independent oscillators the internal and dynamical time arrows are identical. But the coupling between the coordinates may change this situation and flip or destroy certain time arrows. For instance when an oscillator is coupled to coordinates with different internal time arrows its (dynamical) time arrow may simply not exist and the oscillator becomes acausal. Our main interest is to understand the rules the system time arrow is defined by the retarded system Green-function, in particular to find the conditions for the (dynamical) time arrow be set independently of the internal time arrow, the case called irreversibility. Acausality, the nonexistence of dynamical time arrow is a special case of irreversibility.

We return to a single oscillator in {\bf Section \ref{nullsp} Time arrow and the null-space} where its retarded Green-function is written as $D^r=D^n+D^f$ where $D^n(-t)=D^n(t)$ and $D^f(-t)=-D^f(t)$ are called near and far Green-functions. It is easy to see that $D^n$, a particular solution of the inhomogeneous equation of motion produces the response of the oscillator to an external source in a time reversal invariant manner. Hence the contribution to the dynamical time arrow comes from $D^f$ which handles solutions of the homogeneous equation of motion, modes lying in the null-space of the equation of motion operator. The lesson is that we have to trace the way the far environment Green-functions appear in the solution in our search for the origin of the system time arrow.

When the internal time arrows are parallel or opposite for two oscillators then the time arrow is well defined or nonexistent for the coupled oscillators, respectively according to the results of {\bf Section \ref{couplta} Coupling of time arrows}. The state of affairs changes for infinitely many oscillators with continuous spectrum, considered in {\bf Section \ref{conspectr} Continuous spectrum} where it is found that even for parallel time arrow the system may not have a well defined time arrow. In this case the response of the system for an external source is the sum of retarded and advanced terms and the usual causal structure is broken. This happens due to the emergence of poles of the retarded system Green-function at complex frequencies on the non-physical half-plane. The imaginary part of the poles yields a finite life-time for normal modes, ie. irreversibility. 
 
A more detailed, dynamical picture, offered in {\bf Section \ref{ssb} Dynamical breakdown of reversibility and causal structure} is based on an analogy with phase transitions. To motivate this view let us start with a simplified version of ferromagnetic transition, described by a scalar field $\phi(t,\v{x})$ which is subject to the potential $V(\phi)=j\phi+r\phi^2/2+g\phi^4/4$ with $g>0$. The simple mean-field solution, obtained by minimizing the potential predicts the spontaneous breakdown of the discrete symmetry $\phi\to-\phi$ for $j=0$ as a second order phase transition at $r=0$. In the broken symmetry phase with $r<0$ we find a first order phase transition at $j=0$ if the external source $j$ is varied. In fact, the system ''remembers`` the sign of the symmetry breaking term $j$ even after the limit $j\to0$. Thought the symmetry which is broken spontaneously is discrete this first order phase transition is driven by the soft Goldstone modes arising from the breakdown of the continuous symmetry group of the three-space by the presence of domains \cite{tren}.

Phase transitions or spontaneous symmetry breakings take place in infinite systems only and are described in {\bf Section \ref{spsymbr} Spontaneous symmetry breaking} for a realistic, large but finite system as the unexpected emergence of a slow collective coordinate. In case of the spontaneous symmetry breaking the space-average of $\phi(t,\v{x})$, the order parameter displays flip-flops between the minima of the potential in a large but finite system with $r<0$ and the long time average is the symmetric value $\phi=0$. But the typical flip-flop time diverges with the volume and the strictly infinite system develops a non-vanishing order parameter. In case of a phase transition without spontaneous symmetry breaking the role of order parameter can be played by the proportion of the volume of domains of a given phase and the slowing down is experienced without flip-flops which characterize the breakdown of the discrete $Z_2$ symmetry only. Thus phase transitions are characterized by the non-commutativity of the thermodynamical, $N\to\infty$ and the long observation time, $T\to\infty$ limits. The mathematics is simpler when the averages are calculated as
\be\label{nt}
\lim_{N\to\infty}\lim_{T\to\infty}O_N(T),
\ee
called $NT$ limit, because we deal with a finite system and an unlimited precision of observation. The other way of defining averages,
\be\label{tn}
\lim_{T\to\infty}\lim_{N\to\infty}O_N(T),
\ee
the $TN$ limit, may produce surprising results because certain safely looking arguments of the $NT$ procedure fail due to the complications of observing a strictly infinite system within a finite amount of time. Furthermore, in the $NT$ limit all information about the system is recorded since there is enough time to resolve the dynamics of a finite system. We have limited access to the dynamics in the $TN$ procedure where finite time is allowed to observe an infinite system. 

It is important to realize that phase transition and spontaneous symmetry breaking are the results of such a restricted way of observation. The expectation value of the order parameter is vanishing in the ground state or in a statistical description of a symmetrical system, a result which can be proven easily in the finite case, $N<\infty$. But one should not forget that measurements are always performed in finite time period, $T<\infty$ and the large value of $N$ may generate an unexpectedly slow collective mode. For as large $N$ as Avogadro's number some collective coordinate may appear to be static for all practical purposes in a Universe of finite age. If this happens with an order parameter then the symmetry breaking adiabatic approximation applies with excellent precision.

The dynamical breakdown of the time reversal invariance is addressed in {\bf Section \ref{irrev} Irreversibility}. An obvious candidate for the order parameter of irreversibility is the entropy production, $\dot S$. The definition of the entropy is based on limited observability, the separation of the full system into an observed subsystem and its environment where only the former can be irreversible. The time scale of slowing down of the order parameter is the time needed to restore symmetry by ``forgetting'' the initial conditions. This takes place when the entropy becomes saturated by reaching the heat death, $\dot S=0$ and the environment time arrow, set by the intial conditions decouples from the system. The formal analogue of the symmetry breaking external source $j$ for irreversible systems is the $i\epsilon$ term in the retarded Green-functions because the solution of the equation of motion differ by a finite amount when the Green-functions with $\epsilon=0^+$ or $\epsilon=0^-$ are used. But we keep the discussion on a more elementary mechanical level and avoid the use of entropy below.

To exploit the analogy with spontaneous symmetry breaking it is more illuminating to follow the dependence of the solution on the order the thermodynamical and the long observation time limits are carried out. Observations of finite time can not resolve all spectral levels when the spectrum has a condensation point at the lowest system eigenfrequency. As a result, extrapolations based on such limited observations indicate an energy sink in the environment, the dynamical basis of dissipative forces and irreversibility. The necessary condition to generate irreversibility in such a manner is the presence of a condensation point in the environment energy spectrum at a normal frequency of the observed system. One can find similar argument to explain acausality, as well. The time of the action of an external perturbation is reconstructed in {\bf Section \ref{acasch} Acausality} from observations limited in time and acausality is found for sufficiently strong slow soft environment modes.

Such a heuristic approach to irreversibility and acausality calls for a more systematical description of the effective dynamics, constructed for finite time observations. The CTP formalism, used in Sections \ref{cctp}-\ref{qctp} is well suited to achieve this goal. In the CTP method of classical physics, introduced in {\bf Section \ref{cctp} Classical closed time path formalism} the trajectory $x(t)$ for $0<t<T$, subject to the initial conditions $x(0)=\dot x(0)=0$ is extended by flipping the time arrow at $t=T$ and returning the full system to its initial state for $t=2T$, thereby constructing a periodic CTP trajectory $x_{CTP}(t)=x(t)$, $0<t<T$ and $x_{CTP}(t)=x(T-t)$ for $T<t<2T$. It is easier to handle the CTP trajectory by splitting it into two paths,
\be\label{ctptraj}
\begin{pmatrix}x^+(t)\cr x^-(t)\end{pmatrix}=\begin{pmatrix}x_{CTP}(t)\cr x_{CTP}(2T-t)\end{pmatrix}.
\ee
The distinguishing feature of the CTP formalism is the independent handling of the trajectories $x^+(t)$ and $x^-(t)$ that might be interpreted as a reduplication of the degrees of freedom. This enlarges the phase space but the equation of motion closes this gap by imposing the same dynamics for the CTP doublets. Such a redundancy can be stated in a simple manner as a gauge invariance,
\be\label{ctpsym}
x(t)=\frac{1+\kappa(t)}2x^+(t)+\frac{1-\kappa(t)}2x^-(t)
\ee
reflected by the solution of the equation of motion with an arbitrary function $\kappa(t)$. This scheme admits a variational principle for initial condition problems and offers a simple definition of retarded Green-functions even for acausal systems. The reduplication of the degrees of freedom makes even possible to find dissipative Euler-Lagrange equation. Finally, it separates the oriented and non-oriented interactions among degrees of freedom in time thereby offering a clearer insight into the way the environment time arrows influence the system. The action whose variational equation identifies the CTP trajectory is introduced in {\bf Section \ref{ctpacts} Action}. The opposite orientation of time for $t<T$ and $t>T$ suggests \be\label{nctpact}
S_{CTP}[x^+,x^-]=S[x^+]-S[x^-],
\ee
where $S[x]$ denotes the action of the full system. The gauge invariance \eq{ctpsym} should warn us that this action has a large degree of degeneracy. In fact, the action \eq{nctpact} is vanishing for $x^+(t)=x^-(t)$. The simplest way achieving this, together with the regularization of the Green-function with continuous frequency spectrum, is the introduction of an imaginary term in the action similar to Feynman's $i\epsilon$ prescription.

The CTP formalism developed in {\bf Section \ref{ctpcsho} System of harmonic oscillators} solves a riddle about acausality, as well. Acausality poses a challenge because the simple integration of the equations of motion in time with an external source can not produce an acausal result. But the integration of the equations of motion becomes ill-defined for infinite systems due to the possible non-commutativity of the limits $\dt\to0$ and $N\to\infty$. The way out from this ambiguity is the use of the Green-function formalism which provides a well defined scheme to handle infinite systems. The order of limits \eq{tn}, to be followed in this case corresponds to periodic trajectories with period length $2T$ in time. Introducing an external source $j(t)=j_0\delta(t-t')$ coupled to the coordinate $x$ generates a causal response in the CTP trajectory $x_{CTP}(t)$ for $t'\le t\le 2T-t'$. Thus causality appears as a cancellation of the response for $0<t<t'$ and  $2T-t'<t<2T$, a highly non-trivial phenomenon for periodic trajectories. This can be lost in the effective dynamics constructed by limited observations where the flipping of the time arrow at $t=T$ can not be performed for the unresolved slow environment modes. These continue their motion for $t>T$ as far as the effective system equation of motion is concerned. However, the resolved part of the whole system experiences the flipped time arrow for $t>T$ and the cancellation needed for causality is destroyed. 

The quantum version of the model \eq{lagrho} is treated within the quantum CTP scheme in {\bf Section \ref{qctp} Quantum systems}, and the same Green-functions are recovered as in the classical case. The formalism is introduced for quantum systems in {\bf Section \ref{qctpf} CTP formalism} and the ways to recognize the classical limit, quantum fluctuations and decoherence are pointed out. The issue of the orders of the limits \eq{nt},\eq{tn} is taken up again in {\bf Section \ref{explirr} Explicit breakdown of time reversal invariance} because the two orders yield different expectation values even for reversible models. It is Fubini's theorem, applied to Feynman graphs which guarantees the same result for both orders \eq{nt},\eq{tn}. But a necessary condition for this theorem to hold, the absolute convergence of the loop-integrals is violated by the UV divergences of quantum field theories, the natural framework for large quantum systems. One usually follows the order \eq{nt} but the redefinition of the counterterms, the bare parameters of the theory make possible to implement the scheme \eq{tn}, too. When aiming at detecting a possible irreversibility as a phase transition one should not change counterterms at the phase boundary. Rather, one should use a scheme which is applicable to both phases. The strategy to save the scheme \eq{nt} is based on another way of identifying spontaneously broken symmetries instead of following a dynamical procedure, the checking of the slowing down of the order parameter, one carries out a test in equilibrium by introducing an explicitly symmetry breaking term in the action and studies the limit when the strength of this term approaches zero. A scheme which can be used with the usual procedure \eq{nt} for irreversible systems consists of performing the limit $\epsilon\to0$ in the Green-functions with continuous spectrum, corresponding to infinitely long observational time. In fact, a small but finite $\epsilon$ generates a finite life-time and represents a dynamical realization of the restriction to long time observations mentioned in Section \ref{irrev}. The small but finite $\epsilon$ scheme agrees with the usually applied infinitesimal $\epsilon$ prescription in the path integral representation. But a non-Hermitean term introduced in the Hamiltonian generates non-unitary time evolution and breaks the time translation invariance of Green-functions. A simple possibility is proposed which suppresses the non-translation invariant part of the Green-functions, the introduction of a degenerate anti-Hermitian operator in the Hamiltonian which generates irreversibility by means of friction forces.

As mentioned before a characteristic feature of the CTP formalism is that it extends the dynamics over a non-physical regime by the reduplication of the degrees of freedom. This extension is not explorable by the classical trajectory due to the gauge invariance \eq{ctpsym}. Quantum effects, however render them different since $x^+(t)-x^-(t)$ may be considered within the path integral representation as the source of quantum fluctuations. On the level of expectation values, corresponding to the solution of the equation of motion in classical mechanics, one always recovers the equivalence of the two doublers, $\la x^+(t)\ra=\la x^-(t)\ra$ but one can gain an interesting insight into the effects of quantum fluctuations by looking into the interference pattern between the time axes as they build up the observed expectation values. Two such effects, related to unitarity of the time evolution and decoherence are discussed in {\bf Section \ref{undec} Unitarity and decoherence}. It is shown that the system remains reversible and causal as long as its time evolution which may involve memory effects is unitary. Hence the introduction of an infinitesimal imaginary part in the Hamiltonian like the one mentioned above is a necessary modification to recover acausality. Another, special feature of quantum fluctuations is decoherence, the recovery of classical probabilities. It is shown that irreversibility and decoherence are generated by the same dynamical mechanism, both being due to the same spectral strength. Their only difference, discovered in case of mixed initial and final  conditions for the environment is that the soft quantum fluctuations are weighted with their time arrow for irreversibility and without time arrow for the classical limit.

\section{Classical harmonic systems}\label{clho}
A simple toy model, a set of coupled harmonic oscillator is introduced in this section with special attention paid to the time arrow of each oscillator. The first question is the manner the time arrow is sent through the trajectory of a single driven harmonic oscillator from its initial condition. After that the problem of coexistence and competition of different time arrows are considered in the simplest case when two oscillators are coupled. Finally the competition of time arrows is followed for infinitely many oscillators with continuous spectrum.

\subsection{System of oscillators}\label{sysoscs}
The Hamiltonian of the model \eq{lagrho}, written as
\be\label{hamho}
H=\frac{m}{2}\dot x_0^2+\hf\left(m\omega_0^2-\sum_n\frac{g_n^2}{m\omega^2_n}\right)x_0^2-jx_0+\sum_n\left[\frac{m}2\dot x^2_n+\frac{m\omega_n^2}2\left(x_n+\frac{g_nx_0}{m\omega_n^2}\right)^2\right]
\ee
shows that this model has a stable dynamics as long as
\be\label{stab}
\sum_n\frac{g_n^2}{m\omega^2_n}<m\omega_0^2.
\ee

The solution of the equations of motion
\bea\label{thoem}
j&=&m\ddot x_0+m\omega^2_0x_0+\sum_{n=1}^Ng_nx_n,\nn
0&=&m\ddot{x}_n+m\omega_n^2x_n+g_nx_0.
\eea
for the time interval $0\le t\le T$ is rendered unique by suitable boundary conditions in time. The degrees of freedom in the environment are subject of initial or final conditions, $x_j(0)=\dot x_j(0)=0$ or $x_j(T)=\dot x_j(T)=0$ with $j=0,1,\ldots,N$, respectively. The boundary conditions in time define time arrow for the decoupled system, $g_n=0$. The coordinate $x_j$ with initial or final boundary condition is said to have the time arrow $\tau_j=1$ or $\tau_j=-1$, respectively. We use initial conditions for the system, $\tau_0=1$.

We seek the system dynamics. For this end the environment coordinates are expressed by means of the second equation in \eq{thoem} as
\be\label{elim}
x_n(\omega)=\frac{g_n}{m[(\omega+i\epsilon\tau_n)^2-\omega_n^2]}x_0(\omega),
\ee
where the imaginary term in the denominator with $\epsilon=0^+$ takes care of the orientation of time. This expression is inserted into the first equation, resulting the system trajectory
\be\label{hosyssource}
x(t)=\int_0^Tdt'D^r(t,t')j(t'),
\ee
given by the retarded Green-function 
\be
D^r(t,t')=\int_{-\infty}^\infty\frac{d\omega}{2\pi}e^{-i\omega(t-t')}D^r(\omega)
\ee
where
\be\label{thoretprop}
D^r(\omega)=\frac1{m[(\omega+i\epsilon)^2-\omega^2_0]-\Sigma^r(\omega)}
\ee
with the self-energy
\be\label{selfen}
\Sigma^r(\omega)=\sum_{n=1}^N\frac{g^2_n}{m}\frac1{(\omega+i\epsilon\tau_n)^2}.
\ee
The roots or the inverse Green-function, $D^{r-1}(\omega)=0$ belong to the normal mode frequency spectrum.

When the spectrum of the model $N\to\infty$ becomes continuous or at least develops condensation points then it is advantageous to use the spectral functions defined for both time arrows separately,
\be\label{spectrfncd}
\rho_\tau(\Omega)=\sum_{n=1}^N\Theta(\tau\tau_n)\frac{g_n^2}{2m\omega_n}\delta(\omega_n-\Omega),
\ee
satisfying the condition \eq{stab},
\be
\int\frac{d\Omega}\Omega[\rho_+(\Omega)+\rho_-(\Omega)]<\frac{m\omega^2_0}2.
\ee
Models with spectral function $\rho(\Omega)=\ord{\Omega^p}$ are called Ohmic for $p=1$ and dynamics with $p<1$ or $p>1$ are called sub- or super-Ohmic, respectively \cite{weiss}. A simple phenomenological form for an Ohmic system with a smooth Debey-cutoff is
\be\label{spectrfncdd}
\rho_\tau(\Omega)=\Theta(\Omega)\frac{g_\tau^2\Omega}{m\Omega_D(\Omega_D^2+\Omega^2)},
\ee
with $\Omega_D>0$ and $g_+^2+g_-^2<m^2\omega^2_0\Omega^2_D/\pi$. The corresponding self energy,
\be\label{fbselfen}
\Sigma^r(\omega)=\sum_\tau\frac{g_\tau^2}{m\Omega_D}\int_0^\infty d\Omega\frac{\Omega}
{\Omega_D^2+\Omega^2}\frac{2\Omega}{(\omega+i\epsilon\tau)^2-\Omega^2},
\ee
can easily be found,
\be\label{debself}
\Sigma^r(\omega)=-\frac{i\pi}m\sum_\tau\frac{\tau g^2_\tau}{\Omega_D(\omega+i\tau\Omega_D)},
\ee
together with the Green-function
\be
D^r(\omega)=\frac1{m[(\omega+i\epsilon)^2-\omega^2_0]+\frac{i\pi}{m\Omega_D}(\frac{g^2_+}{\omega+i\Omega_D}-\frac{g^2_-}{\omega-i\Omega_D})}.
\ee

It is instructive to compare these results with that corresponding to a simpler spectral function with sharp cutoff $\Lambda$ \cite{caus},
\be\label{spectrfncdds}
\rho'(\Omega)=\Theta(\Omega)\frac{g^2}{m\Omega^2_0}\left(\frac\Omega{\Omega_0}\right)^p.
\ee
We present the formulae for oscillators with forward internal time arrow only for simplicity, where with $2\Lambda^{p+1}g^2<(p+1)m^2\omega^2_0\Omega_0^{p+2}$ which gives for few selected values $p=-1,0,1$ and 3,
\bea
D^r_{-1}&=&\frac1{m(\omega^2-\omega_0^2)+\frac{i2\pi g^2}{m\Omega_0\omega}},\nn
D^r_1&=&\frac1{m(\omega^2-\omega_0^2)+\frac{g^2}{m\Omega_0^3}(2\Lambda+i\pi\omega)},\nn
D^r_3&=&\frac1{m(\omega^2-\omega_0^2)+\frac{g^2}{m\Omega_0^5}(\frac{\Lambda^3}3+\Lambda\omega^2+i\pi\omega^3)},\nn
D^r_0&=&\frac1{m(\omega^2-\omega_0^2)-\frac{g^2}{m\Omega_0^2}(\ln\frac{\omega^2}{c^2\Lambda^2}+i\pi)}.
\eea
The equation of motion of models with super-Ohmic spectrum is a differential equation with higher order derivatives. The equation of motion of sub-Ohmic systems becomes a more involved integro-differential equation with memory kernel. The border line between the two families, the Ohmic case, displays the usual friction force and represents the only causal and dissipative local equation of motion. Another lesson of the comparison of the self-energy of the spectral functions with different cutoffs is that the way the high frequency modes are suppressed may influence the low frequency behavior of the Green-function.

It is sometime advantageous to use the normal modes 
\be
y_\alpha=\sum_{j=0}^NA^{-1}_{\alpha j}x_j,
\ee
where the mixing coefficients satisfy the sum rule $\sum_\alpha A^2_{j\alpha}=1$ and yield the retarded Green-function
\be\label{retnorm}
D^r(\omega)=\sum_\alpha\frac{A^2_{0\alpha}}{m[(\omega+i\epsilon)^2-\omega_\alpha^2]}.
\ee
The normal mode spectral function
\be
\rho_n(\Omega)=\sign(\Omega)\sum_\alpha\delta(\omega_\alpha-|\Omega|)\frac{A^2_{0\alpha}}{2m\omega_\alpha},
\ee
defined in terms of the normal mode frequencies $\omega_\alpha$ allows us to write the retarded Green-function as
\bea\label{dretnormm}
D^r(\omega)&=&\int\frac{d\Omega\rho_n(\Omega)}{\omega+i\epsilon-\Omega},\nn
D^r(t)&=&-i\Theta(t)\int_{-\infty}^\infty d\Omega\rho_n(\Omega)e^{-i\Omega t}.
\eea
Note that whenever there is a condensation point in the spectrum $\{\omega_n\}$ then the normal mode spectrum $\{\Omega_\alpha\}$ displays condensation point, as well.

\subsection{Time arrow and the null-space}\label{nullsp}
Let us now take a single oscillator, corresponding to the system coordinate $x_0$ in our model, decoupled from the environment, $g_n=0$. The simplest way to observe the dynamical time arrow locally, in an arbitrarily small time interval around an observation time $t'$ is to find the response to an external perturbation $j(t)=j_0\delta(t-t')$. The role of the boundary conditions in time is to define what is kept fixed when the external source is varied. The time arrow point in the direction in which the response to the variation of $j_0$ is found. When the response is non-vanishing before and after the perturbation then no time arrow exists. The initial or final conditions are usually taken into account by the choice of an appropriate solution of the homogeneous equation of motion. Hence the time arrow, defined by such test is handled by the null-space of the equation of motion operator $D_0^{-1}=-m(\partial_t^2+\omega_0^2)$, consisting of functions of the form $x_0(t)=x_{01}\cos(\omega_0 t)+x_{02}\sin(\omega_0 t)$, the solution of the source-less oscillator problem. The functions of this null-space are free fields in classical electrodynamics and mass-shell modes in quantum field theories. We shall call modes inside and outside of the null space free and driven modes, respectively. The variational equation of motion determines the driven modes only because the free modes drop out from the Lagrangian \cite{maxwell} and are fixed by the boundary conditions rather than dynamical equations. In fact, the response of the oscillator to an external source is ill defined, it diverges at resonance when the frequency of the source approaches $\omega_0$. 

The usual way of separating the null-space and the rest is the introduction of far and near Green-functions by analogy with electrodynamics,
\bea\label{nfhogffs}
D_0^f(\omega)&=&-\frac{i\pi}m\sign(\omega)\delta(\omega^2-\omega_0^2),\nn
D_0^n(\omega)&=&P\frac1{m(\omega^2-\omega_0^2)},
\eea
where $P$ denotes the principal part. The time-dependent forms are
\bea\label{nfhogf}
D_0^f(t,t')&=&-\frac{\sin\omega_0(t-t')}{2m\omega_0},\nn
D_0^n(t,t')&=&-\frac{\sin\omega_0|t-t'|}{2m\omega_0}.
\eea
The far or near Green-function handle the free and the driven modes, respectively and the retarded and advanced Green-functions are $D_0^{\stackrel{r}{a}}=D_0^n\pm D_0^f$. The driven modes are generated by the given external source but have no time arrow. In fact, an external source $j(t)=j_0\delta(t-t')$ generates a response both for $t<t'$ and $t>t'$ because the near Green-function is symmetric, $D_0^n(t)=D_0^n(-t)$. The time arrow is introduced by the antisymmetric far Green-function, $D_0^f(-t)=-D_0^f(t)$.

The lesson of these trivial remarks is that we have to separate the components of the null-space from the rest in discussing the issue of the time arrow. A time reversal invariant linear equation of motion generates a symmetric response for the external source but leaves the issue of the null-space open. The mathematical source of an eventual anomaly of the time arrow is a strong singularity of the near Green-function at the null-space which manages to mix these two components. Such a mixing takes place when oscillators with different null-spaces are coupled because the non-dynamical free modes of one oscillator are coupled to the dynamical, driven modes of the other. This is the mechanism by which the environment boundary conditions influence the system time arrow.

\subsection{Coupling of time arrows}\label{couplta}
What happens with the time arrows in our model, defined for $g_n=0$ when the oscillators become coupled? The answer is presented for the simplest case of $N=1$, when the retarded system Green-function
\be\label{retgfndc}
D^r(\omega)=\frac1{m[(\omega+i\epsilon)^2-\omega^2_0]-\frac{g_1^2}{m}\frac1{(\omega+i\epsilon\tau_1)^2-\omega_1^2}}
\ee
has four poles, $\omega_{\sigma,\sigma'}=\sigma\omega_{\sigma'}+i\eta_{\sigma'}$ with $\sigma,\sigma'=\pm1$. Due to the stability condition \eq{stab} $\omega_{\sigma'}$ and $\eta_{\sigma'}$ are real and $\eta_{\sigma'}=\ord{\epsilon}$,
\bea
\omega_\pm&=&\sqrt{\hf(\omega^2_0+\omega^2_1\pm d)},\nn
\eta_\pm&=&-\epsilon\begin{cases}1&\tau_1=1\cr\pm\frac{\omega_1^2-\omega_0^2}d&\tau_1=-1\end{cases},
\eea
with $d=\sqrt{(\omega^2_0-\omega^2_1)^2+4g^2/m^2}$. In case of parallel decoupled time arrows the interaction does not change the time arrow as expected. But the environment destroys the system time arrow when the decoupled time arrows are conflicting. When the two null-spaces agree, $\omega_0=\omega_1$ and the time arrows are conflicting then the poles remain on the real axis and the conventional Green-function method fails.

The generalization of these results for larger, finite $N$ can easily be imagined though the details are rather involved. The null-space of the normal modes is $2N$ dimensional and these free modes are fixed by the boundary conditions. The contribution of the normal modes can easily be seen by using the partial fraction decomposition for the Green-function \eq{thoretprop} where each partial fraction corresponds to a normal mode. If all degrees of freedom have forward decoupled time arrows then the stability condition \eq{stab} makes all poles of $D^r_0(\omega-i\epsilon)$ real and there is a global time arrow. In case of conflicting time arrows in the decoupled case the normal modes possess well defined orientation in time in the absence of real poles. But some accidental degeneracy of the decoupled null-spaces with opposite time arrow may create ill posed problem.

What happens in the limit $N\to\infty$? There should be no qualitative change as long as the spectrum remains discrete, without condensation point. The distinguishing feature of the discrete spectrum is that the analytical continuation of Green-functions over the complex frequency space is unique. This property, together with the stability condition \eq{stab} keep the normal frequency spectrum real and the observations, made for finite systems apply. But a condensation point or a continuous part in the spectrum may lead two important changes. When environmental free modes with opposite time arrow converge in frequency towards a system free mode then for strong enough environmental spectral strength the minority system time arrow may simply be erased. This phenomenon, irreversibility, is the subject of the rest of this work.

\subsection{Continuous spectrum}\label{conspectr}
When the environment spectrum has a continuous spectrum then the spectral functions \eq{spectrfncd} yield the retarded system Green-function
\be\label{grfncs}
D^r(\omega)=\frac1{m[(\omega+i\epsilon)^2-\omega_0^2]+\pi\frac{g^2_+\Omega_{D-}(\Omega_{D-}+i\omega)+g^2_-\Omega_{D+}(\Omega_{D+}-i\omega)}{m\Omega_{D+}\Omega_{D-}(\omega+i\Omega_{D+})(\omega-i\Omega_{D-})}}.
\ee
The poles of the self energy, given by Eq. \eq{selfen} merge and accumulate their imaginary parts in Eq. \eq{debself}. This self energy generates poles in Eq. \eq{grfncs} with finite imaginary part when inserted in the Green-function \eq{thoretprop}, indicating that normal modes are damped and have finite life-time, the microscopic manifestation of irreversibility. Note that the Green-function is in general acausal, there are poles on the unphysical sheet. Poles with finite imaginary part, leading to reversibility are present for spectral weights with different suppression for large frequencies suggesting that the breakdown of the time reversal invariance arises from the behavior of the spectral strength at vanishing frequency and reversibility may prevail for spectra with a single condensation point, too.

In the case of a pure initial condition problem, $g_-=0$ the local time arrows, defined for $g_n=0$ agree. Nevertheless the Green-function \eq{grfncs} may be acausal, the condensation point in the spectrum of forward moving modes can generate a backward propagating signal. We shall return to the origin of this rather surprising effect later.

\section{Dynamical breakdown of reversibility and causal structure}\label{ssb}
We set identical, forward time arrow for each oscillator from now on and turn to the question of the breakdown of the time reversal invariance of the effective system dynamics in the thermodynamical limit $N\to\infty$. For finite $N$ the effective system equation of motion, obtained by eliminating the environment coordinates is a differential equation of order $2N$, needs $2N$ boundary conditions and remains reversible. The effective equation of motion continues to be reversible in the limit $N\to\infty$ for discrete spectrum due to the availability of the normal mode decomposition and the Green-function \eq{dretnormm}. But this state of affairs may change when the spectrum has a condensation point and infinitely many poles coalesce as in the Green-function \eq{grfncs} with $g_-=0$. The effective equation of motion needs less boundary conditions then expected and the remaining ''invisible`` boundary conditions are the source of irreversibility. It is argued below that this mechanism is similar to spontaneous symmetry breaking.

\subsection{Spontaneous symmetry breaking}\label{spsymbr}
The Green-function \eq{thoretprop}-\eq{selfen} is unique with well defined analytical continuation over the complex frequency space for discrete spectrum and the order of the summation in the self energy \eq{selfen} and the integration over the frequency in Fourier integrals to find the time dependence of the Green-function is arbitrary. But if there is a condensation point in the spectrum then these operations may not commute anymore. The relevance of the order of the limits where the system size and observation time is sent to infinite is a hallmark of spontaneous symmetry breaking. Note that the normal mode Green-function \eq{dretnormm} remains always causal because the identification of the normal modes always precedes the frequency integration. 

What is the correct order of limits? Mathematical ambiguities always point to a choice in the preparation or in the observation of the system and the choice between the $NT$ and the $TN$ limits, given by Eqs. \eq{nt}-\eq{tn}, respectively depends on the physical circumstances. It is instructive to consider the status of rotational symmetry in a macroscopic body of characteristic length scale $\ell=1cm$, being composed of $N\sim10^{24}$ atoms of mass $m$ whose elementary interactions are assumed to be invariant under translations and rotations. We separate the translational and rotational motion by distinguishing laboratory and body-fixed, co-moving coordinate systems, the latter is defined by having vanishing center of mass velocity and diagonal tensor of inertia
\be
\Theta^{jk}=m\sum_nx^j_nx^k_n.
\ee
Denote the translation and rotation which bring the laboratory frame into the body-fixed coordinate system by $T(\v{X})$ and $R(\theta,\phi,\alpha)$, respectively and use the coordinates $\v{X}$, the Euler angles $\theta,\phi$ and $\alpha$ and  $3N-6$ relative coordinates expressed in the body-fixed frame to describe the positions of the particles of the body. The effective dynamics of the three Euler angles, considered as order parameters for rotation is driven by the Hamiltonian
\be\label{hamrot}
H_{rot}=\hf L^j(\Theta^{-1})^{jk}L^k+\ord{L^4},
\ee
written in terms of the total angular momentum $\v{L}$. 

One can show that the ground state of a finite system described by a single component wave function and obeying the Schr\"odinger equation with regular potential is non-degenerate \cite{feynmannd}. This result makes the angular momentum in the ground state of arbitrary large but finite bodies, held together by rotational invariant forces vanishing and excludes spontaneous symmetry breaking. The resolution of this apparent contradiction is provided by the adiabatic,classical approximation which breaks rotational symmetry for $N\to\infty$ and is an excellent approximation for as large $N$ as Avogadro's number. The Euler angles of a finite body follow classical dynamics in a good approximation because the large inertia tensor makes the rotational excitation spectrum dense. The low lying rotational excitations decouple from the remaining, internal degrees of freedom and the ground state can be constructed by means of angular momentum eigenstates whose number is approximately $N$-independent giving $\la L_j\ra=\hbar\ord{N^0}$. The typical angular velocity is therefore of the order of magnitude
\be
\omega\sim\frac\hbar{Nm\ell^2}\sim10^{-27}~sec^{-1}
\ee
where the mass of the proton is used for $m$. The exceedingly large time for a complete turnaround makes the rotation unobservable and the macroscopic body appears to break rotational symmetry in its ground state.

Note the important role intial conditions play in this mechanism. Spontaneous symmetry breaking takes place when the ground state or the ensemble average display less symmetry than the underlying equations of motion, when the initial conditions prevent the system to explore the phase space prescribed by symmetry. If more energy is available in the initial state then the order parameter may move faster and the symmetry may appear to be restored. In other words, critical slowing down prevents the system to ``forget'' its initial conditions which were imposed at as low energy as possible.

\subsection{Irreversibility}\label{irrev}
Our goal is now to understand irreversibility as a natural result of the widening of the discrete spectrum lines due to insufficient observation time \cite{caus}. Let us return to the question of the issue of the limits $NT$ and $TN$. Observations made according to the $NT$ procedure can resolve every excitation frequency, can trace the energy exchange among coordinates and can recognize the detailed, microscopically reversible dynamics. This is not the case for the $TN$ procedure where no finite observation time can resolve all frequencies around a condensation point. To see this point clearer let us follow the system for a finite amount of time and look into its response to the source $j(t)=j_0\delta(t-t')$. The finiteness of the observation time is taken into account by considering the product
\be\label{obstraj}
x^{obs}_0(t)=c(t)x_0(t)
\ee
as the measured result where $c(t'+t)=c(t'-t)$ is an IR cutoff, $c(t'+t)=1$ for $|t|\ll T$ and $c(t'+t)=0$ for $|tT\gg T$. The observed frequency spectrum
\be
x^{obs}_0(\omega)=\int_{-\infty}^\infty\frac{d\omega'}{2\pi}c(\omega-\omega')D^r(\omega')j(\omega')
\ee
can be written as
\be
x^{obs}_0(\omega)=\int_{-\infty}^\infty\frac{d\Omega\rho^{obs}(\Omega)}{\omega+i\epsilon-\Omega}j_0
\ee
where 
\be
\rho^{obs}(\Omega)=\sum_\alpha\frac{A^2_{0\alpha}}{4\pi m\omega_\alpha}[c(\Omega-\omega_\alpha)-c(\omega_\alpha+\Omega)]
\ee
denotes the spectral density extracted from the limited observations. If there is time to resolve all spectral levels, $c(\omega_\alpha-\omega_\beta)\approx0$ $\alpha\ne\beta$ then 
we find peaks, 
\be
\rho^{obs}(\Omega)\approx\sign(\Omega)\sum_\alpha\frac{A^2_{0\alpha}}{4\pi m\omega_\alpha}c(\Omega-\omega_\alpha)
\ee
whose separation is more obvious by making observations in longer time. But there is always an unresolved part of the spectrum with condensation point and the corresponding infinitely many normal modes provides the energy sink needed for dissipation. An IR cutoff on observations introduces coarse graining and that generates irreversibility \cite{zwanzig}. The spectral lines spread even if the source $j(t)=j_0\delta(t-t'')$ with $t''\ne t'$ is located in an asymmetric manner in time with respect to the IR cutoff except that the speed of spreading depends on $t''-t'$.
 
The characteristic feature of irreversibility, generated by the spontaneous breakdown of the time reversal invariance is that the action is real and poles of the Green-functions with non-vanishing real part appear in pairs, $\omega_\pm=\pm\omega_1+i\omega_2$. The real part, $\pm\omega_1$ is the remnant of the formal time reversal invariance of the original dynamics and the imaginary part, $i\omega_2$ reflects irreversibility.

\subsection{Acausality}\label{acasch}
It was pointed out above that the Green-function \eq{grfncs} is acausal even if $g_-=0$ when initial conditions are imposed for all degrees of freedom. The origin of this surprising effect can also be understood by taking into account the finiteness of observation time. We use now the observed trajectory \eq{obstraj} to reconstruct the source
\be
j^{obs}(\omega)=D^{r-1}(\omega)x^{obs}_0(\omega),
\ee
written as
\be\label{jobs}
j^{obs}(\omega)=\int\frac{d\omega'}{2\pi}c(\omega-\omega')\frac{D^r(\omega')}{D^r(\omega)}j(\omega').
\ee
The partial fraction decomposition can be used to obtain a Green-function \eq{thoretprop}-\eq{selfen} as a sum of simple pole terms,
\be
D^r(\omega)=\sum_p\frac{Z_p}{\omega-\omega_p}.
\ee
The contribution of the term $p$ dominates the integral in Eq. \eq{jobs} for $\omega'\sim\omega_p$,
\be
j^{obs}_p(\omega)\approx\int\frac{d\omega'}{2\pi}c(\omega-\omega')\frac{\omega-\omega_p}{\omega'-\omega_p}j(\omega').
\ee
We choose the IR cutoff $c(\omega)=2\eta/(\omega^2+\eta^2)$ with $\eta=2\pi/T$ and find for the source $j(t)=j_0\delta(t)$
\be
j_p^{obs}(t)=j_0\left[\delta(t)-\frac{2\pi}Te^{-i\omega_pt-2\pi\frac{|t|}T}\right].
\ee
Apparent acausality results from the difficulties in reconstructing a sharply localized source in time due to the abundant soft modes with $\omega_p\approx0$.

\section{Classical closed time path formalism}\label{cctp}
It was pointed out in Section \ref{nullsp} that the driven modes build up an unoriented motion in time and the time arrow is set by the free modes. It is easy to recast the equation of motion in a slightly modified form where these two classes of modes are clearly separated. Apart of making this point explicitly, this formalism is needed, as well, to tackle difficulties of infinite systems. The point is that the solution of the equations of motion of a finite system can easily be obtained by direct integration in time for $N<\infty$ but complications arise when spontaneous symmetry breaking is suspected for $N\to\infty$ because both exact solutions and the numerical quadratures belong to the $NT$ scheme. A simple formalism presented in this Section allows us to treat infinite systems by means of powerful functional methods (Green-functions) in either the $NT$ or in the $TN$ scheme.

One needs the inverse of differential operators when a degree of freedom is eliminated by its equation of motion. It is a well known problem that the self-adjoint extension of the derivative operator $id/dt$ requires well defined boundary conditions, not available in an initial condition problem. How to save the functional method in this case? The existence of Green-functions follows from the functional setting of the action principle. In fact, let us take a dynamical system described by the coordinate $x$ and the action,
\be\label{clact}
S[x]=\hf xD^{-1}x+\ord{x^3},
\ee
written in condensed notation, introduce an infinitesimal external source $j(t)$ for diagnostic goal and define the functional
\be\label{clw}
W[j]=S[x]+jx
\ee
in which the trajectory $x$ is chosen to satisfy the variational equation of motion,
\be\label{clem}
\fd{S[x]}{x}+j=0.
\ee
The form
\be\label{clwev}
W[j]=-\hf jDj+\ord{j^3},
\ee
identifies the Green-function as a second functional derivative,
\be\label{clgf}
D=\fdd{W[j]}{j}{j}_{|j=0}.
\ee

This procedure can not handle initial conditions in time because the quadratic form $D^{-1}$ of the action must be symmetric, $D=D^n$, excluding the null-space modes in the solution of the equation of motion. Another face of this problem is that the variational equation must be imposed at the end point $t=T$ if an initial condition problem is considered and this condition cancels the final momentum. To avoid this restriction we flip the time arrow at the end of the motion and return the system backward in time to its initial state. It turns out that such an extension of the motion leads to non-symmetrical Green-functions, such as $D^r$ in the variational equations. The trajectory $x_{CTP}(t)$, $0<t<2T$ obtained in this manner can be broken into a pair of trajectories with different time orientations \cite{maxwell} as in Eq. \eq{ctptraj} and this formalism goes over to the CTP method \cite{schw} when considered in quantum mechanics.

\subsection{Action}\label{ctpacts}
There are two different realizations of the CTP action principle, one for $T<\infty$ and another for $T=\infty$. The limit of $T\to\infty$ of the first is not continuous, it does not produce the scheme $T=\infty$ because the former has no null-space in the quadratic form of the linear equation of motion. Let us start with the scheme $T<\infty$ by introducing two trajectories for each degree of freedom, $x(t)\to(x^+(t),x^-(t))=\hx(t)$, satisfying the boundary conditions $x^\pm(0)=0$, $\dot x^\pm(0)=v_i$, $x^+(T)=x^-(T)$ and the classical CTP action is defined as
\be\label{ctpact}
S_{CTP}[\hx]=\int_0^Tdt[L_{CTP}(x^+(t),\dot x^+(t))-L_{CTP}^*(x^-(t),\dot x^-(t))],
\ee
$\Re L_{CTP}=L$ being the original Lagrangian of the model. The minus sign in the action reflects the opposite orientation of time for the two CTP trajectories. When the original, real Lagrangian is used then the CTP action is vanishing whenever $x^+(t)=x^-(t)$. This degeneracy is due to the lack of initial condition for the velocity. Such an initial condition could in principle be imposed by using the extended action $S[\hx]\to S[\hx]+S_{bc}[\hx]$ with
\be\label{lagrmult}
S_{bc}[\hx]=\ell\left(2p_i-\frac{\partial\Re L[x^+]}{\partial\dot x}_{|t=0}-\frac{\partial\Re L[x^]-}{\partial\dot x}_{|t=0}\right)
\ee
where $\ell$ is a Lagrange multiplier and $p_i$ denotes the initial generalized momentum. A simplified scheme is used below with $v_i=0$ corresponding to $\ell=0$ for theories with space inversion invariance. The degeneracy of the CTP action for real Lagrangian and $\ell=0$ for $x^+(t)=x^-(t)$ can easily be lifted by making a small imaginary shift of the harmonic potential,
\be\label{ctplag}
L_{CTP}(x,\dot x)=L(x,\dot x)+i\frac{m\epsilon}2x^2
\ee
for a particle of mass $m$ and the limit $\epsilon\to0$ is performed after deriving and solving the variational equations of motion. One could have used real shift of the potential but the the imaginary shift is closer to the way the degeneracy is lifted for $T=\infty$. 

Let us consider the simplest example, a harmonic oscillator in the presence of a driving source $j$ when the CTP action can be written as
\be\label{ctpactho}
S_{CTP}[\hx]=\int dt\left[\hf\hx(t)\hD^{-1}\hx(t)+\hj(t)\hx(t)\right]
\ee
with $\hj=(j,-j)$. The equation of motion $\hx=-\hD\hj$ for the physical trajectory is
\be\label{ctpsolho}
x(t)=-\sum_{\sigma'}\int_0^Tdt'D^{\sigma\sigma'}(t,t')\sigma'j(t'),
\ee
for $0\le t\le T$ where $D^{\sigma\sigma'}=(\hD)^{\sigma\sigma'}$ denotes the matrix elements of the CTP Green-function with $\sigma,\sigma'=\pm1$ and $\sigma$ is arbitrary. The independence of the trajectory \eq{ctpsolho} of the choice of $\sigma$ yields the CTP relation
\be\label{ctprel}
D^{++}+D^{--}=D^{-+}+D^{+-},
\ee
assuring that the Green-function can be parameterized by three functions,
\be\label{ctpgrfncf}
\hD=\begin{pmatrix}\bar D+D^n&\bar D-D^f\cr \bar D+D^f&\bar D-D^n\end{pmatrix}.
\ee
The inverse Green-function $\hD^{-1}$ is a symmetric operator, $\hD^{-1tr}=\hD^{-1}$ hence so is $\hD$ and $\bar D(t,t')=\bar D(t',t)$, $D^n(t,t')=D^n(t',t)$ and  $D^f(t,t')=-D^f(t',t)$. According to Eq. \eq{ctpsolho} the retarded Green-function is $D^r=G^n+G^f$ and the advanced Green-function will be defined by $D^a=D^n-D^f$. The function $\bar D(t,t')$ can not be identified by the help of Eq. \eq{ctpsolho}. The explicit calculation of $D^n$ and $D^f$ is given in Appendix \ref{finitetgrfv}. 

The limit $T\to\infty$ of the CTP Green-function is discontinuous because the coupling of the two time axis, introduced at $t=T$, disappears if $T=\infty$ and the Green-function recovers translation invariance in time. How to maintain the coupling between the two time axis, the distinguished feature of the CTP scheme? This coupling is supposed to handle the null-space modes of the linearized equation of motion hence one expects the form
\be
S_{CTP}[\hx]=\int_{-\infty}^\infty dtL(x^+,\dot x^+(t))-\int_{-\infty}^\infty dtL(x^-,\dot x^-(t))+S_{BC}[\hx]
\ee
where a translation invariant $S_{BC}[\hx]=\ord{\hx^2}$ singles out the null-modes which couple the two time axis. The determination of the Green-function starts with the observation that the Fourier transform of $D^n$ and $D^f$ defined by Eq. \eq{ctpgrfncf} are real and imaginary, respectively. This property, together with the identity
\be
\frac1{x+i\epsilon}=P\frac1x-i\pi\delta(x)
\ee
requires the usual form, Eqs. \eq{nfhogffs}, for $D^n(\omega)$ and $D^f(\omega)$. A simple form of $\hD$ which reproduces the desired $D^n$ and $D^f$ is
\be\label{ctpgrfvf}
\hD(t,t')=-\frac{i}{2m\omega_0}\begin{pmatrix}e^{-i\omega_0|t-t'|}&e^{-i\omega_0(t-t')}\cr e^{i\omega_0(t-t')}&e^{i\omega_0|t+t'|}\end{pmatrix},
\ee
and a formal prescription to derive it as $T\to\infty$ is presented in Appendix \ref{infinitetgrfv}. We are interested in a trajectory for $t\sim T/2$, safely away from the endpoints $t=0$ and $t=T$ which should decouple from the dynamics to recover translation invariance in time as $T\to\infty$. In order this to happen we would need short enough life-time, $\omega_0/\epsilon\ll T$, generated by the imaginary term in the action for the pole contributions. However such a removal of the dependence on the end points leads to decoupling of the two time axis taking place at $t=T$ and thus the CTP scheme is lost. Therefore we have to rely on the cancellation due to the fast rotating phase of the pole contributions as $T\to\infty$ to recover translation invariance in time. The usual way to realize this scheme is to treat $\epsilon$ as a formal, infinitesimal quantity. 

The discontinuity of the Green-functions in reaching the continuous spectrum inherent in setting $T=\infty$, is reminiscent of phase transitions and in this respect the continuous spectrum and $T$ correspond to the adiabatic approximation and the system size, respectively. The realistic observations belong to the scheme of $T<\infty$, containing the scale parameter $T$. If $T$ is well beyond the observation scales then it is more natural to work out formal rules for the case of continuous spectrum where this scale parameter is absent, and to use them to describe observations made at large enough $T$.

The inverse of the Green-function, calculated in Appendix \ref{invgrfv},
\be\label{inctpgrfv}
\hD_0^{-1}(\omega)=m\left[(\omega^2-\omega_0^2)\begin{pmatrix}1&0\cr0&-1\end{pmatrix}
+i\epsilon\begin{pmatrix}1&-2\Theta(-\omega)\cr-2\Theta(\omega)&1\end{pmatrix}\right],
\ee
is the quadratic part of the translation invariant CTP action for $T=\infty$. It shows that the coupling between the time axis interchanges modes with given time orientation and the time arrow of the solution is set by the "interference" between the time axes. The inverse Fourier transform of this expression yields
\be\label{coupledtas}
S_{BC}[\hx]=\frac\epsilon\pi P\int_{-\infty}^\infty dtdt'\frac{x^+(t)x^-(t')}{t-t'}+\frac{i\epsilon}2\int_{-\infty}^\infty dt[x^{+}(t)-x^{-}(t)]^2,
\ee
where the recognition of the time orientation of the modes requires a non-local term. The coupling of finite strength between the time axis at $t=T$ of the scheme $T<\infty$ is traded into an infinitesimally strong coupling in the latter, smeared out over an infinitely long time evolution. 

We have two different ways of introducing the intial conditions. One possibility is to set up initial conditions at some initial time $t_i\sim T/2$ by means of Lagrange multipliers. For such a trajectory the integral on the right hand side of Eq. \eq{ctpsolho} extends between $t_i$ and infinity and $D^n+D^f$ with continuous frequency spectrum is just the retarded Green-function as expected. Another possibility is to start the motion with $x_i=\dot x_i=0$ at $t=0$ and let the source $j$ drive the trajectory to the desired initial conditions used before at $t_i\sim T/2$. As the source $j$ varies for $0<t<t_i$ $x(t)$ and $\dot x(t)$ sweep through the set of accessible initial conditions. For such initial conditions the two realizations of the motion are equivalent and we can use the simpler scheme, starting the motion at $t_i$ and using translation invariant Green-functions. By enlarging the class of dynamical quantities which enter in the Lagrangian with an external source we can always describe the actual experimental preparation of the system and render our intial conditions accessible. This is the way the restriction $v_i=0$ on the initial condition, mentioned after Eq. \eq{lagrmult} can be avoided in the CTP formalism.

\subsection{System of harmonic oscillators}\label{ctpcsho}
The steps presented in Section \eq{sysoscs} to find the system trajectory can be repeated within the CTP formalism without difficulty. Eqs. \eq{thoem} then refer to CTP pair of trajectories, the inverse system Green-function is of the form
\be\label{ctpgf}
\hD^{-1}=\hD^{-1}_0-\hat\sigma\hat\Sigma\hat\sigma
\ee
where the $\hD_0^{-1}$ is given by Eq. \eq{inctpgrfv}, the matrix
\be
\hat\sigma=\begin{pmatrix}1&0\cr0&-1\end{pmatrix}
\ee
represents the different signs in front of the two actions in Eq. \eq{ctpact} in the coupling between the oscillators. The CTP matrix of the self-energy,
\be
\hat\Sigma(\omega)=\sum_{n=1}^Ng^2_n\hD_n(\omega),
\ee
contains $\hD_n(\omega)$, the Green-function of the $n$-th environment oscillator which is calculated in Eq. \eq{ctppropfr} for $n=0$. The spectral representation for the self-energy is
\be\label{spectrrse}
\hat\Sigma(\omega)=\sum_{\tau=\pm1}\int_0^\infty d\Omega2\Omega\rho_\tau(\Omega)\hD_0(\tau\omega,\Omega),
\ee
$\hD(\omega,\Omega)$ being defined in Eq. \eq{ctppropfr} for $T\to\infty$. The off-diagonal elements in \eq{ctppropfr} can be used to find the useful relations
\bea\label{spfninv}
\bar\Sigma(\omega)&=&-\pi[\rho_+(|\omega|)+\rho_-(|\omega|)],\nn
\Sigma^f(\omega)&=&-i\pi\sign(\omega)[\rho_+(|\omega|)-\rho_-(|\omega|)].
\eea

It is an important property of the CTP Green-functions that $\hD$, $\hat\sigma\hD^{-1}_0\hat\sigma$ and $\hat\Sigma$ all have the block structure shown in Eq. \eq{ctpgrfncf}, allowing to define the functions $D^n$, $D^{-1n}$, $\Sigma^n$, etc. One carries out the spectral integral first in the self energy within the scheme \eq{tn} and the limit $T\to\infty$ follows after that. The spectral weight \eq{spectrfncdd} results
\bea
\Sigma^n(\omega)&=&-\frac\pi{m}\sum_{\tau=\pm1}\frac{g^2_\tau}{\Omega^2_{D\tau}+\omega^2},\nn
\Sigma^f(\omega)&=&-i\frac{\pi\omega}{m}\sum_{\tau=\pm1}\frac{g^2_\tau\tau}{\Omega_{D\tau}(\omega^2+\Omega_{D\tau}^2)},\nn
\Sigma^i(\omega)&=&-i\frac{\pi|\omega|}{m}\sum_{\tau=\pm1}\frac{g^2_\tau}{\Omega_{D\tau}(\omega^2+\Omega_{D\tau}^2)}.
\eea
The inversion in Eq. \eq{ctpgf} can easily be carried out \cite{ed} and one finds 
\bea\label{retgrfv}
D^n&=&D^rD^{-1n}D^a\nn
D^f&=&-D^rD^{-1f}D^a\nn
D^i&=&-D^rD^{-1i}D^a
\eea
in particular
\be
D^{\stackrel{r}{a}}=\frac1{D_0^{-1\stackrel{r}{a}}-\Pi^{\stackrel{r}{a}}},
\ee
in agreement with Eq. \eq{thoretprop} and the results of Section \ref{sysoscs} are reproduced within the CTP formalism. Despite this equivalence the CTP formalism offers advantages when effective equations of motion are dissipative. In fact, it is known that variation method can not lead to non-conservative dynamics. But the reduplication of degrees of freedom offers a possibility to derive from the action principle, cf. the odd powers of $\omega$ in the denominator of the Green-function \eq{grfncs}.

Another advantage is that this formalism offers a mechanism explaining the loss of causality for dissipative systems, noted in Section \ref{ssb}. The price of transforming an initial condition into a variational problem is to make both the trajectories and external perturbations periodic in time. The response on the trajectory $x_{CTP}(t)$, used in Eq. \eq{ctptraj} to a perturbation made at time $t'$ is restricted to the time interval $t'<t<2T-t'$. Thus causality becomes the result of a rather fragile cancellation within the time interval $2T-t'<t<2T$ when the equation of motion is integrated from $t=0$ to $t=2T$ and inaccuracies like the one mentioned in Section \ref{acasch} may lead to acausal dynamics.

One may get an insight into such a breakdown of causality by looking into the ``polarization cloud`` the system generates in the environment, the correlation between the system and the environment motion. This correlation represents the ''dressing`` of the system particle by the environment, the emergence of a system quasi-particle. When the time arrow of the effective system dynamics is flipped at time $T$ in the framework of the CTP formalism then the motion of the environment which is related to the system is reversed, as well. But irreversibility, the spontaneous breakdown of time reversal invariance implies a limited resolution of the soft slow environmental modes by the system. Such a limited resolution leads to an incomplete reversal of the environment motion at $t=T$ within the effective CTP dynamics. In other words, that part of the ''polarization cloud`` which belongs to the very soft modes and remain unresolved within the limited observation time continues its motion and what is turned back at $t=T$ is actually only the observed fraction of the dressed quasi-particle. It is natural that this deformed quasi-particle follows a path backwards in time which is different from the forward motion and the cancellation for $2T-t'<t<2T$ is incomplete.

\section{Quantum systems}\label{qctp}
The system of quantum harmonic oscillators, defined by the Hamiltonian \eq{hamho} can be solved in a manner analogous to the discussion of Section \ref{sysoscs}. The equations of motion \eq{thoem} are imposed on the coordinate operators taken in the Heisenberg representation and the solutions can be written by means of the same retarded Green-function as in the classical case leaving the result of Sections \ref{clho}-\ref{ssb} valid for quantum oscillators. Nevertheless one gains more insight and efficiency for interactive, anharmonic models when the CTP formalism is followed in the path integral representation.

\subsection{CTP formalism}\label{qctpf}
The quantum mechanical expectation value
\be\label{expval}
\la\psi(t)|A|\psi(t)\ra=\la\psi(0)|U^\dagger(t,0)AU(t,0)|\psi(0)\ra,
\ee
where $|\psi(0)\ra$ denotes the initial state and $U(t,0)$ stands for the time evolution operator. It is crucial to realize that one has to treat the operators $U^\dagger(t,0)$ and $U(t,0)$ separately in the expectation value to build up eventual interactions in perturbation expansion and to generate the observable $A$ by functional derivatives. Such an independent treatment is possible if the source $j$ is allowed to be different in the two operators. Such a non-physical, formal extension of the formulae of the expectation value not only allows us to derive Green-functions for such an initial condition problem but gives access to the density matrix, as well. The basic idea is therefore to deal with two time axes in an explicit manner, one for the operator $U$ and another for $U^\dagger$. Such a point of view proposed in the original CTP formalism \cite{schw}, based on the generator functional
\be\label{gfqctp}
e^{\ih W[\hj]}=\Tr[U(j^+;T,0)\rho(0)U^\dagger(-j^-;T,0)].
\ee
where $U(j;T,0)$ is the time evolution operator in presence of the external source $j$ coupled linearly to the coordinate and $\rho(0)$ denotes the initial density matrix. The physical system with unitary time evolution corresponds to the choice $j^+=-j^-$. It is not difficult to find the path integral representation for this functional,
\be\label{ctppint}
e^{\ih W[\hj]}=\int D[\hx]e^{\ih S_{CTP}[\hx]+\ih\int dt\hj(t)\hx(t)},
\ee
where the action is defined by Eqs. \eq{ctpact} and \eq{ctplag} and the integration is over trajectories with fixed initial points, $x^\pm(0)=x_i$ and arbitrary common final point, $x_f=x^+(T)=x^-(T)$. 

The expectation value of the coordinate is given by the functional derivative,
\be
\la\psi(t)|A|\psi(t)\ra=\fd{W[\hj]}{j^+(t)}=\fd{W[\hj]}{j^-(t)}.
\ee
The equation of motion for the expectation values belonging to the quantum mechanical intial condition problem, can be derived from an action principle, just as in the classical case. One performs a functional Legendre transform by defining the effective action,
\be\label{legtr}
\Gamma[\hx]=W[\hj]-\hx\hj
\ee
in condensed notation where
\be\label{legtrv}
\hx=\fd{W[\hj]}{\hj}.
\ee
The inverse Legendre transformation is given by \eq{legtr} and the change of variable
\be
-\hj=\fd{\Gamma[\hx]}{\hx},
\ee
whose defining equation is interpreted as an equation of motion. Note that Eqs. \eq{clact}-\eq{clem} define similar Legendre transformation in classical mechanics, leading back to the classical action as effective action. Naturally the effective action is different from the original one when some environmental degrees of freedom are eliminated, and effective dynamics can be considered both in classical and quantum cases.

It is instructive to extend the CTP formalism to an Open Time Path (OTP) scheme, based on the generator functional
\be
e^{\ih W[\hj;x_f^\pm]}=\la x^+_f|U(j^+;T,0)\rho(0)U^\dagger(-j^-;T,0)|x^-_f\ra
\ee
to be interpreted as the matrix element of the density operator $\la x^+_f|\rho(T;\hj)|x^-_f\ra$, subject to a generalized time evolution which has external source $j^+$ and $j^-$ for $U$ and $U^\dagger$, respectively. The corresponding path integral expression, \eq{ctppint} contains open trajectories, with common initial point, $x^\pm(0)=x_i$ and independent final points, $x^\pm(T)=x^\pm_f$. Note that the integrand of the path integral is the product of factors, one depending on $x^+$ and the other one on $x^-$ if the initial state is pure, the initial density matrix is factorisable and all degrees of freedom are present. The two trajectories of the pair $\hx$ are uncorrelated in the path integral and the OTP formalism produces in this case simply the product of a transition amplitude and its complex conjugate in one expression. 

The formalism with double time axes becomes advantageous when mixed initial state is considered or some degree of freedom is eliminated because the intuitive scheme of Feynman diagrams can be maintained to represent expectation values in terms of elementary processes. Though this issue is not important when harmonic systems are considered there is another advantage which is relevant here. The integrand of the path integral is not factorisable when mixed states contribute to the expectation value and this makes the trajectories of the pair correlated within the path integral. We can follow the quantum-classical transition by inspecting such correlations generated by the CTP action
\be
S_{CTP}[\hx]=S[x^+]-S^*[x^-]+S_i[\hx],
\ee
which is not additive due to the influence functional \cite{feynmanv} $S_i[\hx]$. The integrand of the usual path integral with a single time axis is the contribution of a given trajectory to a transition amplitude \cite{feynmanh}. In a similar manner that integrand of the OTP path integral represents the contribution of the pair of trajectories, $\hx$ to the density matrix. Its suppression for $x^+_f\ne x^-_f$ signals decoherence in the coordinate basis \cite{zeh,zurek}. When all degrees of freedom are present then the integrand is a pure phase and its suppression may come from fast rotating phase as the trajectories are varied. When degrees of freedom have already been eliminated then $\Im S_i[\hx]\ge0$ represents environment induced decoherence. 

The two trajectories of the pair are always identical, $x^+(t)=x^-(t)$ for the solution of the equation of motions in the classical CTP formalism. The solution of the classical equation of motion corresponds to expectation value in the quantum case and the time axes give identical time evolution for the expectation value of the coordinate. But as soon as we look into the interference pattern between the time axes quantum fluctuations appear in $x^+(t)-x^-(t)$ \cite{caldeiraqle} witnessing incomplete decoherence and indicating that the possibility of separating the two trajectories of the CTP pair is an $\ord{\hbar}$ genuine quantum feature.

\subsection{Explicit breakdown of time reversal invariance}\label{explirr}
We make a little digression in this Section and consider a technical problem in the procedure of detecting irreversibility, generated by the well known divergences of quantum field theory. The question of the order of the limits $NT$-$TN$ is specially acute for the perturbation series of interactive quantum systems where it corresponds to the choice of the order of summation over the energies and the momenta of intermediate states. In fact, Fubini's theorem guarantees the independence from the order of energy and momentum integration for finite Feynman graphs only. Divergent graphs change their values when the order of integration is changed. One usually follows the $NT$ scheme and integrates over energies first because this scheme allows the proper treatment of infrared divergences when necessary. One may use the scheme \eq{tn} as soon as the thermodynamical limit is proven to be safe, the price being a modification of the counterterms controlling the ultraviolet dynamics. But rather than making the change $NT\leftrightarrow TN$ at a phase transition it is more preferable to rely on the same $NT$ procedure in exploring the whole phase diagram. Therefore we can not avoid the problem of establishing spontaneously broken symmetries within this scheme. The usual solution, mentioned in Section \ref{syn} is based on equilibrium observables, one breaks the symmetry by an external source $j$, coupled to the order parameter. Since the symmetry is broken explicitly for a fixed $j\ne0$ the order parameter assumes a constant, nontrivial value and the scheme $NT$ is obviously applicable. The symmetry is recovered or remains broken spontaneously after the limit $j\to0$ has been performed in the symmetric and the broken symmetry phase, respectively. 

The simplest way to implement this procedure for irreversible systems is to keep the $i\epsilon$ term in the denominators of the Green-functions small but finite rather than infinitesimal. The imaginary part of the poles of the Green-functions spreads the discrete spectrum lines and the resulting finite life-time of excitations represents a dynamical realization of the IR cutoff for observations, introduced by hand in Section \ref{irrev}. 
Now we have three limits to perform, $N\to\infty$, $T\to\infty$ and $\epsilon\to0$. Since the $i\epsilon$ term regulates the pole contributions to the Green-functions the limit $\epsilon\to0$ must be carried out after the other two limits. One can define in this manner the limits $\epsilon NT$,
\be\label{ent}
\lim_{\epsilon\to0}\lim_{N\to\infty}\lim_{T\to\infty}O_{\epsilon N}(T),
\ee
and $\epsilon TN$,
\be\label{etn}
\lim_{\epsilon\to0}\lim_{T\to\infty}\lim_{N\to\infty}O_{\epsilon N}(T),
\ee
and either of them can be used for irreversible systems.

It should be noted that neither of the limits $\epsilon NT$ or $\epsilon TN$ is equivalent with the usual implementation of Feynman's $i\epsilon$ prescription because $\epsilon$ is not infinitesimal. To see the difference between small but finite and infinitesimal $\epsilon$ let us briefly outline a slight generalization of the procedures $\epsilon NT$ or $\epsilon TN$ where irreversibility is achieved by introducing an $\ord{\epsilon}$ non-Hermitean term in the Hamiltonian $H=H(t)-iK$ with $H^\dagger(t)=H(t)$, $K^\dagger=K$  where $H(t)$ might be time dependent due to some external sources. It is the non-unitarity of the time evolution which introduces complications in setting up the Heisenberg representation where the equations of motion are easier to access. Let us start with the Schr\"odinger representation where operators $A_S$ are time independent and states are subject of the Schr\"odinger equation $i\hbar\partial_t|\psi(t)\ra_S=(H(t)-iK)|\psi(t)\ra_S$. One can define a diffusive Heisenberg representation by performing the basis transformation 
\bea
|\psi(t)\ra_d&=&U^\dagger(t,t_i)|\psi_t\ra_S,\nn
A_d(t)&=&U^\dagger(t,t_i)A_SU(t,t_i)
\eea
with
\be
U(t,t')=T[e^{-\ih\int_{t_i}^tdt'H(t')}]
\ee
which places the unitary time dependence in the operators and generates the equations of motion
\bea
i\hbar\partial_t|\psi(t)\ra_d&=&-iK_d(t)|\psi(t)\ra_d\nn
i\hbar\partial_tA_d(t)&=&[A_d(t),H_d(t)].
\eea
The expectation values satisfy the equation of motion,
\be\label{eqmdh}
\partial_t\la\psi(t)|A_d(t)|\psi(t)\ra=\frac1{i\hbar}\la\psi(t)|[A_d(t),H_d(t)]-i\{A_d(t),K_d(t)\}|\psi(t)\ra
\ee
showing an unexpected effect of non-unitary time evolution, the loss of invariance under translation in time for Green-functions. This happens because the factors of the time evolution operators corresponding to non-Hermitean Hamiltonian in the usual Heisenberg representation do not simplify in the Green-functions according to the usual rule $U(t_{n+1},t_i)U^\dagger(t_n,t_i)=U(t_{n+1},t_n)$. A more physical way to see this is to note that the introduction of a non-Hermitean part in the Hamiltonian is a dynamical way to implement an IR-cutoff like the one used in in Section \ref{irrev} because $1/\epsilon$ is proportional to the life-time of the states. Since all state decay the initial time when the initial state with a given norm is set remains always explicitly present in the expectation values. Irreversibility is generated even in the scheme $\epsilon NT$ as long as $\epsilon\ne0$ because the integration contour of the Green-functions runs at finite distance from the poles what can be interpreted as a ''natural``, finite width of the spectral lines.

It is instructive to find the analogue of Ehrenfest's theorem in this representation. For this end we choose the Hamiltonian
\be
H(t)=\frac{p^2}{2m}+V(x)-j(t)x,
\ee
and use $K=\hf\hbar\epsilon$. The equations of motion
\bea
\frac{d}{dt}\la x\ra&=&\frac{\la p\ra}m-\epsilon\la x\ra,\nn
\frac{d}{dt}\la p\ra&=&\la j-V'(x)\ra-\epsilon\la p\ra,
\eea
can be written as a Newton equation with a friction term,
\be\label{newtonfrict}
m\frac{d^2}{dt^2}\la x\ra=\la\openone\ra j-\la V'_{eff}(x)\ra-2\epsilon m\frac{d}{dt}\la x\ra
\ee
with an effective potential
\be\label{effpot}
V_{eff}(x)=V(x)+\hf m\epsilon^2x^2,
\ee
reflecting a suppression of deviation from the initial position by the decay of the state norm.

\subsection{Unitarity and decoherence}\label{undec}
Let us return to our simple harmonic model defined by the Lagrangian \eq{lagrho} where the effective system dynamics is described by the the generator functional
\bea\label{quwfu}
e^{\ih W[\hj]}&=&\int D[\hat x]e^{\frac{i}{2\hbar}\hx\hD^{-1}\hx+\ih\hj\hx}\nn
&=&e^{-\frac{i}{2\hbar}\hj\hD\hj},
\eea
of the form identical to the classical counterpart shown in Eq. \eq{clwev}. It gives rise to the Green-function \eq{clgf},
\bea
\fdd{\ih W[\hj]}{\ih j^+(t)}{\ih j^+(t')}_{|\hj=0}
&=&\la0|T[x(t)x(t')]|0\ra=i\hbar D^{++}(t,t'),\nn
\fdd{\ih W[\hj]}{\ih j^-(t)}{\ih j^-(t')}_{|\hj=0}
&=&\la0|T[x(t')x(t)|0\ra^*=i\hbar D^{--}(t,t'),\nn
\fdd{\ih W[\hj]}{\ih j^+(t)}{\ih j^-(t')}_{|\hj=0}&=&\la0|x(t')x(t)0\ra=i\hbar D^{+-}(t,t'),\nn
\fdd{\ih W[\hj]}{\ih j^-(t)}{\ih j^+(t')}_{|\hj=0}&=&\la0|x(t)x(t')0\ra=i\hbar D^{-+}(t,t'),
\eea
defined by following the convention of including some $i\hbar$ factors. Due to the identity of the functional $W[\hj]$ for classical and quantum harmonic systems, Eqs. \eq{clwev} and
\eq{quwfu}, respectively, the classical and quantum CTP Green-functions agree. The Legendre transformation \eq{legtr}-\eq{legtrv} leads back the the classical action for harmonic system and the argument of Section \ref{ctpcsho} applies for the quantum harmonic oscillators, as well.

We now consider two genuine quantum issues, the role of unitarity of the time evolution and of the classical limit. First it is pointed out that unitarity protects both reversibility and causality of the dynamics. In fact, both irreversibility and acausality are found to be generated by the modified pole structure of the Green-functions, the migration of poles to complex location or appearance of new poles. The necessity of poles with non-vanishing imaginary part to generate irreversibility indicates that this latter needs non-unitary time evolution. A dissipative force, like the friction force in Eq. \eq{newtonfrict} appears because the large number of environmental soft modes mix with the system state and makes it ``leaking'' to the environment. As of causality is concerned, note first that the final time $T$ of the time evolution in the generating functional \eq{gfqctp} can be chosen in an arbitrary manner as long as it exceeds the time of any observation. In fact, the extension of the time interval the system is followed, $T\to T+T_1$ amounts to the insertion of the operator $U^\dagger(j^+;T+T_1,T)U(-j^-;T+T_1,T)$ within the trace on the right hand side of Eq. \eq{gfqctp}. But this operator is the identity for unitary time evolution, $j^-=-j^+$ and does not change the generator functional. In other words, any modification of the physical external source $j=j^+=-j^-$ is unobservable by looking into expectation values at times before the modification, briefly causality can not be lost by a unitary time evolution. 

Decoherence, a necessary conditions for classical limit stands for the suppression of the contribution to the CTP path integral for well separated pairs of trajectories. It is advantageous to introduce the Keldysh parameterization \cite{keldysh} of the coordinates, $x=(x^++x^-)/2$, $x^a=x^+-x^-$ and write the $\ord{\hx^2}$ part of the action in Eq. \eq{quwfu} as
\be
\hx\hD^{-1}\hx=\hf xD^{-1a}x^a+\hf x^aD^{-1r}x+\hf x^a\bar D^{-1}x^a,
\ee
where $D^{-1r}$, $D^{-1a}$ and $\bar D^{-1}$ are defined as in Eq. \eq{ctpgrfncf} for $\hat\sigma\hD^{-1}\hat\sigma$, $\hD^{-1}$ being given by Eqs. \eq{inctpgrfv}-\eq{ctpgf}. The quadratic forms $D^{-1r}(t,t')$ and $D^{-1a}(t,t')$ are real but $\bar D^{-1}(t,t')$ is imaginary. Hence the magnitude of the contributions to the path integral \eq{quwfu} is controlled by the last term. Due to the positive definiteness $\Im\bar D^{-1}(\omega)>0$, $\bar D^{-1}(t,t')$ is the origin of environment induced decoherence.

It was pointed out in Section \ref{nullsp} that the time arrow of a driven harmonic oscillator is set by the homogeneous solution of the equation of motion. In a similar manner we can write the retarded Green-function of the effective theory as a sum of a symmetric and asymmetric component, $D^r=D^n+D^f$, the latter, $D^f$ being responsible of the homogeneous solution and the orientation of the time arrow. Eqs. \eq{spfninv} and \eq{retgrfv} show that the time arrow is now set by the combination of the spectral functions $\rho_+(|\omega|)-\rho_-(|\omega|)$ and the suppression is controlled by $\rho_+(|\omega|)+\rho_-(|\omega|)$. The latter combination leads to decoherence and  the former shows the competition between oscillators with oppositely rearranged boundary conditions in time in defining the system time arrow. If this combination is non-vanishing then it dominates the infinitesimal $D_0^f$, given by Eqs. \eq{inctpgrfvc} and may generate irreversibility. The lesson is that for a usual initial condition problem where $\rho_0=0$ irreversibility and decoherence stem from the same dynamical mechanism. Their difference can be explored mathematically by imposing final rather than initial conditions for some environmental degrees of freedom because the irreversibility and decoherence are constructed in the same manner except the time arrow of the environment modes are kept or ignored, respectively. An extreme manifestation of this difference, reversible decoherence is found when the forward or backward moving environment modes have identical spectral weight.

\section{Summary}\label{sum}
The dynamical generation of time arrow of an oscillator is studied in this work within a harmonic model. An oscillator has an internal time arrow, defined by its boundary conditions in time, the description of what is supposed to be kept fixed when its dynamical environment is changing. The influence of the environment at the normal frequency of the oscillator, described by the far Green-function may change the internal time arrow in setting up a dynamical time arrow. Irreversibility corresponds to a complete overwriting of the internal time arrow. In cases when the time arrow is lost as a result of this process one finds acausality.

It is proposed that irreversibility and acausality build up in a manner similar to a phase transition. In particular, irreversibility may be considered as a dynamical symmetry breaking. This analogy is motivated by the slowing down of the order parameter in the thermodynamical limit, the discontinuity of the long time observation limit for infinite systems. Such a mathematical subtlety is important if one keeps in mind that observations in physics always deal with finite systems and infinity or certain limits seem to be useful approximations rather than part of physical reality around us. One may say that the continuous spectrum approximation which allows us to derive irreversibility or even the construction of continuum in mathematics by means of the Bolzano-Weierstrass theorem is an ingenious approximation which eliminates unreasonable large or small quantities, decoupled from the phenomena of interest. 

The particular mechanism to drive a spontaneous symmetry breaking can easily be modeled phenomenologically by introducing an infrared cutoff on the observations. When the dynamics of an infinite system is followed for finite amount of time then the extracted laws may reflect less symmetry. In case the frequency spectrum of our harmonic model possesses a condensation point with sufficiently strong spectral weight the normal modes around the condensation point serve as a sink for the energy received from an external source and observations, carried out during finite amount of time signal irreversibility.

One can always find an order parameter for a spontaneous symmetry breaking. If thermodynamical quantities can be defined then the entropy production is a natural order parameter for irreversibility. Entropy is based on limited observability, the separation of the full system into an observed subsystem and its environment and only the former can be irreversible. This remark opens the possibility that different splitting of the full system into observed and ignored parts lead to different status of the time reversal invariance for the observed component. Since dynamical breakdown of symmetries requires the cooperation of infinitely many degrees of freedom it is reasonable to expect that the ignored component must contain infinitely many sufficiently decorrelated degrees of freedom to show irreversible system dynamics. An interesting problem, left open is to find the ''irreversible universality class``, the subsets of the full system which display irreversibility. 

Despite the similarities between irreversibility and spontaneous symmetry breaking these two phenomena show important differences, as well. On one hand, the conventional spontaneous symmetry breaking is driven by decorrelated domains in three-space. On the other hand, irreversibility happens in time. The dimensionality and the signature of the manifold in the relativistically invariant length differ, as well. The dynamical quantities remain correlated for arbitrarily large time separation, manifested by conservation laws. Further important non-local effects in time are generated in the effective system dynamics by the elimination of degrees of freedom. As a result spontaneous symmetry breaking is made possible along the time, in a one-dimensional chain, the less correlated, ''disordered`` phase belonging to irreversible dynamics. A less formal difference is related to initial conditions. Though the common dynamical origin of the slowing down of the order parameter and irreversibility is the high frequency level density at a system normal frequency these phenomena require different initial conditions. In fact, spontaneous symmetry breaking is observed when the energy in the initial state is as  low as possible. But irreversibility requires some initial energy distributed unequally among the degrees of freedom or the presence of a driving force. The model \eq{lagrho} with weak couplings belong to the latter case because the full system starts close to its minimal energy state and energy can be gained from the source $j(t)$ only. The energy is received by the system and is dissipated to the environment. For sufficiently many soft modes such a dissipation can be maintained for long enough time and the system serves as a transit station of energy between the external source and the environment. It is obviously this energy flux which is the dynamical origin of irreversibility. There are furthermore aspects, such a locality or local equilibrium which are important for phase transitions and can not even be addressed in the model \eq{lagrho}.

The need of handling infinitely many degrees of freedom to establish irreversibility forces the use of more sophisticated methods of solving the equations of motion than their direct integration in time which produces trajectories with unspecified analytical properties. The algebraic elimination of the environment degrees of freedom in Fourier space requires the inversion of some operators which is possible only if their domain, the functional space where they act is properly specified. The mathematical background for inversion is easiest to provide within the action principle. Thus the CTP formalism is worked out for classical systems where the initial condition problem is transformed into a variational problem for periodic trajectories with well defined Green-functions. The near and far Green-functions are constructed for a finite observation time and are calculated explicitly in the long time limit where time translation invariance can be recovered. 

The reduplication of the degrees of freedom, arising from the periodicity of the trajectories in time is an inherent property of the CTP formalism and it explains a surprising feature, the necessity of going beyond the physical space of trajectories to establish an action principle for initial conditions which is local in time. This enlargement of the space of trajectories has no observable effect in classical mechanics where the equations of motion force the CTP doublers to follow the same trajectory. But quantum effects can be encoded in the difference between the doublers owing to the linear superposition principle, since the doublers experience independent quantum fluctuations. One may even define quantum fluctuations in the path integral formalism by the difference of the doublers trajectories. In the classical limit when the system becomes increasingly decohered the doubler trajectories tend to stick together. When the full system starts in thermal equilibrium, represented by a canonical density matrix then there is a specially clear way of separating thermal and quantum fluctuations as fluctuations in the sum and difference of the doubler trajectories in the path integral formalism.  

Decoherence and irreversibility are supposed to be two necessary ingredients of the classical limit. It is found that both are determined by the same spectral functions the only difference being that the former is fed by the unoriented and the latter by the oriented combination of the spectral functions.

The advantage of the simple harmonic model considered in this work is its exact solvability. On the other hand, obvious difficulties hinder the generalization of the results to more realistic anharmonic models with true interactions. Nevertheless, there are good reasons to expect the qualitative results to remain valid for interactive systems, as well. Neither irreversibility nor acausality can be generated in any finite order of the perturbation series. Rather, they must appear on the quasi-particle level of an interactive many-body system. Our harmonic model correspond to quasi-particles obtained by a partial resummation of the perturbation expansion and thus is supposed to be a qualitatively acceptable approximation except in the vicinity of the reversible-irreversible transition. It appears desirable to check if the generalization to interactive systems, where new time scales arise, confirm our general conclusions. A further, more principal question is if the onset of irreversibility is indeed a bona fide phase transition. On one hand, it seems to be the simplest realization of the idea of spontaneous symmetry breaking where the initial conditions modify the long time dynamics by the slowing down of the order parameter. On the other hand, however spontaneous symmetry breaking is a phenomenon where the ground state (respectively the equilibrium state) displays less symmetry than the microscopic equations of motion. This is here not the case because the true ground state (or equilibrium state) is static by nature and thus does not possess a time arrow. Another, related question is if there exist local parameters of the dynamics whose variation drives the system across the reversible-irreversible transition line. This question can not even be addressed within our simple model since locality asks for a field theory. A free field model would be not sufficient either due to its strongly constrained dynamics. Therefore we are led to the problem of including interactions in a field theoretical setting and to search for a nontrivial phase structure with an irreversible phase.

Another issue raised and left open in this work is the different manner acausality may manifest itself. As usual in nonrelativistic physics we define causality as the violation of the order of action and reaction. The question arises if the same mechanism would also lead to a spontaneous breakdown of symmetry with respect to Lorentz boosts in relativistic models.

\acknowledgments
It is pleasure to thank Janos Hajdu for a number of illuminating discussions.

\appendix
\section{CTP Green-function}\label{grfvapp}
The formulae for the Green-function of the CTP formalism are collected in this Appendix. 
First the action is defined for fixed IR and UV cutoffs in Section \ref{ctpactreg}, followed by the derivation of the general expression of the Green-function in Section \ref{ctpgrfvg}. The explicit formulae of the Green-function for finite and infinite time are presented in Sections \ref{finitetgrfv} and \ref{infinitetgrfv}, respectively. Section \ref{invgrfv} is for the calculation of the inverse Green-function, the quadratic form of the kinetic energy in the infinite time case.

\subsection{Action}\label{ctpactreg}
The identification of the end point of the trajectories $x+$ and $x^-$ requires care and we consider a harmonic oscillator with mass $m$ and frequency $\omega_0$ for discrete time, $t=j\dt$, $j=0,1,\ldots,N_{UV}$, $\dt=T/N_{UV}$. The real part of the CTP action is
\be\label{reghoa}
\Re S_{CTP}=\sum_\sigma\sigma\left[\frac{m}{2\dt}\sum_t(x^\sigma_{t+\dt}-x^\sigma_t)^2-\frac{\dt m\omega_0^2}2\sum_tx^{\sigma2}_t\right],
\ee
where the sum over t is for $0\le t\le(N_{UV}-1)\dt$. Partial integration/summation in the kinetic energy yields the form
\bea
\Re S_{CTP}&=&\sum_\sigma\sigma\biggl[\frac{m}{2\dt}\sum_tx^\sigma_t(2x^\sigma_t-x^\sigma_{t-\dt}-x^\sigma_{t+\dt})+\frac{m}{2\dt}[x^\sigma_{N\dt}(x^\sigma_{N\dt}-x^\sigma_{(N-1)\dt})-x^\sigma_0(x^\sigma_\dt-x^\sigma_0)]\nn
&&-\frac{\dt m\omega_0^2}2\sum_tx^{\sigma2}_t\biggr].
\eea
The constraint
\be
S_C=\ell\left[2p_i-\frac{m}\dt(x^+_\dt+x^-_\dt)\right]
\ee
handling the initial condition gives the extended action
\be\label{hoactdet}
S_{CTP}=\hf\sum_{\sigma,\sigma'}\sum_{t,t'}x^\sigma_tD^{-1\sigma,\sigma'}_{0~t,t'}x^{\sigma'}_{t'}+\sum_\sigma\sum_tx^\sigma_t(A^\sigma_t+B^\sigma_ty)
\ee
where the notation $y=x_{N\dt}$ has been introduced and
\bea
D^{-1\sigma,\sigma'}_{d~t,t'}&=&-\delta^{\sigma,\sigma'}\left[\sigma\left(\frac{m}{\dt}\Delta_{t,t'}+\dt m\omega_0^2\delta_{t,t'}\right)-i\dt m\epsilon\delta_{t,t'}\right],\nn
A^\sigma_t&=&-\delta_{t,\dt}\frac{\ell m}\dt,\nn
B^\sigma_t&=&-\delta_{t,T-\dt}\frac{\sigma m}{2\dt}.
\eea
This expression contains an UV and an IR cutoff, $\dt$ and $T$. The UV cutoff is removed first in the continuum limit, $\dt\to0$ by keeping $T$ fixed. The way the IR cutoff is treated is determined by our choice of the scheme $NT$ or $TN$. The removal of the cutoffs is nontrivial because neither of the limits $\dt\to0$ or $T\to\infty$ is continuous.

\subsection{Green-function}\label{ctpgrfvg}
To find the Green-function for the quadratic action \eq{ctpactho}, \eq{hoactdet} we write
\be
S_{CTP}=\hf\hx\hD_d^{-1}\hx+\hx(\hat A+\hat Bz+\hj).
\ee
We keep $\dt$ finite which allows to make the trajectories $x^+(t)$ and $x^-(t)$, given for $0\le t<T$ independent by treating the final point, $z=x^\pm(T)$ separately. Due to the absence of $\ord{z^2}$ term in the action we eliminate first the trajectory $\hx$ by means of its equation of motion, 
\be
\hx=-\hD_d(\hat A+\hat Bz+\hj)
\ee
and express the action in terms of the variable $z$ only,
\be
S_{CTP}=-\hf(\hat A+z\hat B+\hj)\hD_d(\hat A+\hat Bz+\hj).
\ee
The variational equation for $z$,
\be
z=-\frac{\hat B\hD_d(\hat A+\hj)}{\hat B\hD_d\hat B},
\ee
gives finally the action as a functional of the source,
\be
W[\hj]=-\hf(\hat A+\hj)\hD_0(\hat A+\hj),
\ee
by means of the free Green-function
\be\label{ctpgrfnexp}
\hD_0=\hD_d-\hD_d\hat B\frac1{\hat B\hD_d\hat B}\hat B\hD_d.
\ee
The initial condition can be imposed on the trajectory
\be
\hat x=\fd{W[\hj]}{\hj}=-\hD_0(\hat A+\hj)
\ee
by the appropriate choice of the multiplicative factor $\ell$ in $\hat A$.

\subsection{Finite $T$}\label{finitetgrfv}
To find the inverse of $\hD_0^{-1}$ for Eq. \eq{ctpgrfnexp} we have to specify the functional space where we are working. It is span by functions $x(t)$ defined over the time interval $[0,T]$ and satisfying the boundary conditions $x(0)=x(T)=0$,
\be\label{funcspace}
x(t)=\sum_{n=1}^{N_{UV}}\tilde x_n\sqrt{\frac2{T}}\sin\omega_n t,
\ee
with $\omega_n=n\pi/T$ and non-uniform convergence is allowed at $t=T$ in the continuum limit, $N_{UV}\to\infty$ to reach an arbitrary end point $z=\lim_{t\to T}x(t)$. 

The starting point is the Green-function
\be
D_d(t,t')=\frac2{Tm}\sum_{n=1}^{N_{UV}}\frac{\sin\omega_nt\sin\omega_nt'}{\hat\omega^2_n-\omega_0^2+i\epsilon}
\ee
where $\hat\omega_n=\frac2\dt\sin\pi\frac{\dt n}{2T}$. The continuum limit has to be taken with some care. The usual procedure is to split the right hand side into the sum for $1\le n<cN_{UV}$ and for $cN_{UV}\le n<N_{UV}$ where $c$ is a small, $N_{UV}$-independent number. The latter sum, being proportional to $1/N_{UV}c^2$ can be neglected and the replacement $\hat\omega_n\to\omega_n$ can be made in the denominator and we get the diagonal CTP block-matrix
\be\label{ctpgrfndiag}
\hD_d=\begin{pmatrix}D_d&0\cr0&-D_d^*\end{pmatrix}
\ee
with
\be\label{propstal}
D_d(t,t')=\frac2{Tm}\sum_{n=1}^\infty\frac{\sin\omega_nt\sin\omega_nt'}{\omega^2_n-\omega_0^2+i\epsilon}
\ee
after the removal of the UV cutoff.

We need in the denominator in the right hand side of Eq. \eq{ctpgrfnexp}
\be
\frac1{\dt^2}D_d(T-\dt,T-\dt)=\frac{2}{Tm\dt^2}\sum_{n=1}^{N_{UV}}\frac{\sin\omega^2_n(T-\dt)}{\hat\omega^2_n-\omega_0^2+i\epsilon},
\ee
that can be split into the sum of an $\ord{N_{UV}}$ linearly divergent part and a finite piece,
\be\label{dttdtt}
\frac1{\dt^2}D_d(T-\dt,T-\dt)=\frac2{mT}\lim_{N_{UV}\to\infty}\sum_{n=1}^{N_{UV}}\left(1-\sin^2\frac{\dt\omega_n}2\right)+\frac{2\omega_0^2}{mT}\sum_{n=1}^\infty\frac1{\omega^2_n-\omega_0^2+i\epsilon},
\ee
in the continuum limit. Another ingredient of Eq. \eq{ctpgrfnexp} becomes in the limit $N_{UV}\to\infty$
\be\label{ttdt}
\frac1\dt D_d(t,T-\dt)=\frac2{Tm}\sum_{n=1}^\infty\frac{\omega_n\sin\omega_n(t-T)}{\hat\omega^2_n-\omega_0^2+i\epsilon}.
\ee
The complete Green-function \eq{ctpgrfnexp},
\be
\hD_0(t,t')=\begin{pmatrix}D_d(t,t')&0\cr0&-D^*_d(t,t')\end{pmatrix}-\frac{im}{2z\omega_0^2\dt^2}\begin{pmatrix}D_d(t,T-\dt)D_d(T-\dt,t')&D_d(t,T-\dt)D^*_d(T-\dt,t')\cr D^*_d(t,T-\dt)D^+_d(T-\dt,t')&D^*_d(t,T-\dt)D^*_d(T-\dt,t')\end{pmatrix}
\ee
with
\be\label{zctpgrfv}
z=\frac1{T}\sum_{n=1}^\infty\frac\epsilon{(\omega^2_n-\Omega^2)^2+\epsilon^2}.
\ee
converges in the continuum limit.

\subsection{Diverging $T$}\label{infinitetgrfv}
The Green-function can directly be determined for $T=\infty$ as argued in Section \ref{ctpacts} where it was noted that the limit $T\to\infty$ is discontinuous. Our goal here is to find the modification of the rules of the limit $T\to\infty$ to recover the result for $T=\infty$.

First we execute the limit $T\to\infty$ for fixed small but finite $\epsilon$. The diagonal Green-function, $D_d$, given by \eq{propstal} is
\be
D_d(t,t')=-\frac{i}{2m\omega_0}(e^{-i\tilde\omega_0|t-t'|}-e^{-i\tilde\omega_0(t+t')}).
\ee
with $\tilde\omega_0=\omega_0-i\epsilon/2\omega_0$. The expression \eq{dttdtt} becomes
\be
\frac1{\dt^2}D_{0T-\dt,T-\dt}=\frac1{m\dt}-i\frac{\omega_0}{m}
\ee
in the limit $\dt\to0$. The expression \eq{ttdt} requires more care in order to correspond to the motion between the time $t$ and $T$. One finds 
\be
\frac1\dt D_0(t,T-dt)=-\frac1{m}e^{-i\tilde\omega_0(T-t)}
\ee
by keeping $t-T<0$ fixed in the limit $T\to\infty$. We expect translation invariance to recover far from both the initial and the final time hence we seek the Green-function by making the replacement $t\to T/2+t$ with fixed $t$ as $T\to\infty$. The substitution of these results into eq. \eq{ctpgrfnexp} yields
\bea
\hD_0(t,t')&=&-\frac{i}{2m\omega_0}\biggl\{\begin{pmatrix}e^{-i\omega_0|t-t'|}&0\cr0&e^{i\omega_0|t-t'|}\end{pmatrix}\nn
&&+e^{-\frac{\epsilon T}{2\omega_0}}\left[\begin{pmatrix}e^{i\omega_0(t+t'-T)}&e^{i\omega_0(t-t')}\cr e^{-i\omega_0(t-t')}&e^{-i\omega_0(t+t'-T)}\end{pmatrix}-\begin{pmatrix}e^{-i\omega_0(t+t')}&0\cr0&e^{i\omega_0(t+t')}\end{pmatrix}\right]\biggr\}.
\eea

A simple form, given by Eq. \eq{ctpgrfvf} reproduces Eqs. \eq{nfhogf} and yields
\be
\bar D_0(t)=-i\frac{\cos\omega_0 t}{2m\omega_0}.
\ee
This Green-function \eq{ctpgrfvf} can be recovered if the limit $T\to\infty$ is performed by first approximating the infinite sums in Eqs. \eq{propstal}, \eq{dttdtt}, \eq{ttdt} and \eq{zctpgrfv} for large but finite $T$ by integrals and after that by performing the limits $\epsilon\to0$ and $T\to\infty$ in such joint manner that the limits
\bea\label{epslims}
e^{(i\omega_0-\frac\epsilon{2\omega_0})T}&\to&0,\nn
e^{-\frac\epsilon{2\omega_0}T}&\to&1
\eea
apply within the integrand of the right hand side of \eq{hosyssource}. The two steps contradict because $\epsilon T\to0$ is required by the second limit in \eq{epslims}. However the infinite sum can not be approximated by a contour integral if the distance of the pole from the real axes is smaller than the discrete increment of the summation variable. In other words, the continuous spectrum formalism can not be obtained as the infinitely long time limit. The formal solution is based on contour integrals where an infinitesimal $\epsilon$ is introduced which is strong enough for the first limit in \eq{epslims} and weak enough for the second. This is in agreement with the practice of using $i\epsilon$ as a prescription to avoid poles rather than a finite quantity.

\subsection{Inverse Green-function}\label{invgrfv}
The translation invariant Green-function \eq{ctpgrfvf} can be used to find the CTP action for $T=\infty$. Its Fourier transform,
\be\label{ctppropfr}
\hD_0(\omega)=\frac1m\begin{pmatrix}\frac1{\omega^2-\omega_0^2+i\epsilon}&-2\pi i\Theta(-\omega)\delta(\omega^2-\omega^2_0)\cr
-2\pi i\Theta(\omega)\delta(\omega^2-\omega^2_0)&-\frac1{\omega^2-\omega_0^2-i\epsilon}\end{pmatrix},
\ee
shows the structure mentioned in Section \ref{ctpacts}, namely the trajectory \eq{ctpsolho} is constructed by collecting the unoriented driven modes by $D^n(\omega)$ from one time axis and the free modes are taken from the other time axis by $D^f(\omega)$ in such a manner that the sum is a retarded solution.

To find the inverse of $\hD_0(\omega)$ we use $\delta(x)=\epsilon/\pi(x^2+\epsilon^2)$ for the regulated Dirac-delta. The trivial algebraic inversion yields the form \eq{inctpgrfv} or
\bea\label{inctpgrfvc}
D^{-1n}_0&=&m(\omega^2-\omega_0^2),\nn
D^{-1f}_0&=&i\sign(k^0)\epsilon,\nn
D^{-1i}_0&=&\epsilon.
\eea
The coupling of the two trajectories which is at the final point on the original action \eq{ctpact} is hidden in the limit $T\to\infty$ and is replaced by an infinitesimally strong coupling distributed along the whole time axis in a uniform, translation invariant manner.

\end{document}